\documentclass[aps, prl, twocolumn, superscriptaddress, showpacs]{revtex4}
\usepackage{graphicx}
\usepackage{dcolumn}
\usepackage{bm}
\usepackage{natbib}
\usepackage{upgreek}
\usepackage{amsmath}
\usepackage{amssymb}
\usepackage{color}

\newcommand*{\cm}[1]{#1~cm$^{-1}$}
\newcommand*{\Sm}{SmFe$_3$(BO$_3$)$_4$}
\newcommand*{\Nd}{NdFe$_3$(BO$_3$)$_4$}

\begin{document}

\title{Switching of magnons by electric and magnetic fields in multiferroic borates}

\author{A. M. Kuzmenko}
\affiliation{Prokhorov General Physics Institute, Russian Academy of
Sciences, 119991 Moscow, Russia}
\author{D. Szaller}
\author{Th. Kain}
\author{V. Dziom}
\author{L. Weymann}
\author{A. Shuvaev}
\author{Anna Pimenov}
\affiliation{Institute of Solid State Physics, Vienna University of
Technology, 1040 Vienna, Austria}
\author{A. A. Mukhin}
\author{V. Yu. Ivanov}
\affiliation{Prokhorov General Physics Institute, Russian Academy of
Sciences, 119991 Moscow, Russia}
\author{I. A. Gudim}
\author{L. N. Bezmaternykh}
\thanks{deceased}
\affiliation{L. V. Kirensky Institute of Physics Siberian Branch of RAS,
660036 Krasnoyarsk, Russia}

\author{A. Pimenov}
\affiliation{Institute of Solid State Physics, Vienna University of
Technology, 1040 Vienna, Austria}

\begin{abstract}

Electric manipulation of magnetic properties is a key problem of materials research. To fulfil the requirements of modern electronics, these processes must be shifted to high frequencies. In multiferroic materials this may be achieved by electric and magnetic control of their fundamental excitations. Here we identify magnetic vibrations in multiferroic iron-borates which are simultaneously sensitive to external electric and magnetic fields. Nearly 100\,\% modulation of the terahertz radiation in an external field is demonstrated for \Sm. High sensitivity can be explained by a modification of the spin orientation which controls the excitation conditions in multiferroic borates. These experiments demonstrate the possibility to alter terahertz magnetic properties of materials independently by external electric and magnetic fields.

\end{abstract}

\date{\today}


\maketitle

The continuous development of electronic devices drives the necessity to obtain an electric control of magnetic effects~\cite{vaz_jpcm_2012, matsukura_natnano_2015}. Compared to external magnetic field, electric voltage may be applied to smaller spatial area and with much less switching power, thereby improving the performance and increasing the density of integrated components.
In recent years a substantial contribution to achieving electric manipulation of magnetic properties~\cite{matsukura_natnano_2015} has been realized through the application of multiferroics (i.e.,
materials with simultaneous electric and magnetic ordering)\cite{smolenskii_ufn_1982, fiebig_jpd_2005, dong_advph_2015, tokura_rpp_2014, fiebig_natrev_2016}.
In several multiferroics, the coupling between electricity and magnetism is strong enough to allow a mutual influence of both properties. This magnetoelectric coupling has been demonstrated to lead to manipulation of magnetic moments~\cite{lottermoser_nature_2004, chu_natmat_2008, saito_nmat_2009, tokunaga_natmat_2009, choi_prl_2010, oh_natcomm_2014, soda_prb_2016} and magnetic structure~\cite{yamasaki_prl_2007, murakawa_prl_2009, finger_prb_2010, babkevich_prb_2012} by external electric field. These effects have been shown to survive up to room temperature~\cite{zhao_nmat_2006, evans_natcomm_2013}. Recent reviews of the topic can be found in Refs.~[\onlinecite{wang_advph_2009, vaz_jpcm_2012, matsukura_natnano_2015}].

Having in mind possible applications, the time scale of switching is an important issue. For example, in typical ferroelectric devices, this time is limited by the speed of domain wall propagation which sensitively depends upon the amplitude of electric field~\cite{fatuzzo_pr_1962} and may be as short as few tenths of nanoseconds~\cite{tybell_prl_2002, dawber_rmp_2005, grigoriev_prl_2006, ehara_scirep_2017}. In multiferroics the problem of fast switching is not fully settled. Due to low static electric polarization in spin-driven multiferroics~\cite{cheong_nmat_2007, tokura_rpp_2014} substantial degradation of the switching time has been reported~\cite{hoffmann_prb_2011}.
Extremely short switching times of electric polarization and of magnetization can be reached using pulsed laser light. Depending on the specific mechanism of the interaction of the light pulse and the spins, the switching rate may be as short as 40\,fs~\cite{kirilyuk_rmp_2010}. Several interesting recent developments in the field of light-matter interaction include spin modulation via thermalisation processes~\cite{matsubara_natcomm_2015}, pumping the energy into the electronic transitions~\cite{johnson_prl_2012}, using magnetic component of a terahertz pulse~\cite{kampfrath_natph_2011}, or directly exciting the magneto-electric excitation in a multiferroic material~\cite{kubacka_science_2014}. Detailed discussion of the experiment and theory of the short-time optical manipulation of magnetism is given in Refs.~[\onlinecite{kimel_jpcm_2007, kirilyuk_rmp_2010, lambert_science_2014}].

Besides the electric modification of static magnetic structures, a control of the high-frequency properties is of substantial interest~\cite{tokura_phil_2011}. To accomplish this control in the practice, the dynamic processes, which are sensitive to the influence of the static electric field, have to be identified. Especially for terahertz light, the multiferroics are promising as they possess magnetoelectric excitations allowing the combination of electric and magnetic fields. These excitations are called electromagnons~\cite{pimenov_nphys_2006, sushkov_prl_2007, kida_josab_2009} and an external magnetic field may easily control them. However, until now, only a few experiments could demonstrate the electric control of excitations in multiferroics~\cite{rovillain_nmat_2010, shuvaev_prl_2013}. Similar to static experiments, the control here is achieved by modifying the electric domain structure with the gate voltage.
In addition, ferromagnetic resonance in ferromagnetic thin films has been demonstrated to be sensitive to static voltage~\cite{shastry_prb_2004, das_advmat_2009, zhu_prl_2012, nozaki_natphys_2012}. The mechanism of the last effect is generally attributed to the voltage control of the magnetic anisotropy. In this work, we utilize another route to electric control of dynamic magnetic properties based on an influence of electric and magnetic fields on the spin orientation which determines the excitation conditions of fundamental magnetic modes.

Rare-earth iron-borates represent one exotic class of multiferroics~\cite{vasiliev_ltp_2006, kadomtseva_ltp_2010, kadomtseva_jetp_2012}. At high temperatures, all rare-earth borates reveal a non-centrosymmetric trigonal structure belonging to the space group R32~\cite{campa_chemm_1997, hinatsu_jssch_2002, fausti_prb_2006, popova_jre_2009}  which persist down to lowest temperatures for Sm- and Nd-iron-borates\cite{suppl}.
The connection between magnetic and electric ordering in iron borates is realized via the coupling of electric polarization to the antiferromagnetically ordered spin lattice\cite{zvezdin_jetpl_2006, popov_jetp_2010, mukhin_jetpl_2011, ritter_jpcm_2012, popov_prb_2013}.

Without losing any generality, we consider \Sm~ below. As an approximation, the magnetoelectric coupling in iron-borates with easy-plane antiferromagnetic order may be written in the symmetry-dictated form~\cite{zvezdin_jetpl_2006, popov_jetp_2010, mukhin_jetpl_2011}
\begin{equation}\label{eqPL}
  P_x \sim L_x^2 - L_y^2 \ .
\end{equation}
Here, $P_x$ is the electric polarization along the crystallographic $a$-axis and $L_x=M_{1x}-M_{2x}$ and $L_y=M_{1y}-M_{2y}$ are the $x,y$ components of the antiferromagnetic vector of the ordered iron moments. Here, the magnetic structure is modeled by two antiferromagnetically coupled sublattices, $\mathbf{M}_1$ and $\mathbf{M}_2$, respectively (bold symbols denote vectorial quantities).
A peculiarity of Eq.~(\ref{eqPL}) is due to the fact that \Sm~ is an easy plane antiferromagnet. We note that in high enough magnetic fields the antiferromagnetic vector realigns perpendicular to the field (i.e. $\mathbf{L} \perp \mathbf{H}$). In agreement with Eq.~(\ref{eqPL}), for $\mathbf{H}\|b$-axis one obtains~\cite{popov_jetp_2010} $L_x \neq 0, L_y = 0, P_x > 0$, and for $\mathbf{H}\|a$-axis, $L_x = 0, L_y \neq 0, P_x < 0$. That is, the electric polarization rotates by $180^\circ$ after a $90^\circ$ rotation of the external magnetic field.

\begin{figure}[tbp]
\begin{center}
\includegraphics[width=0.99\linewidth, clip]{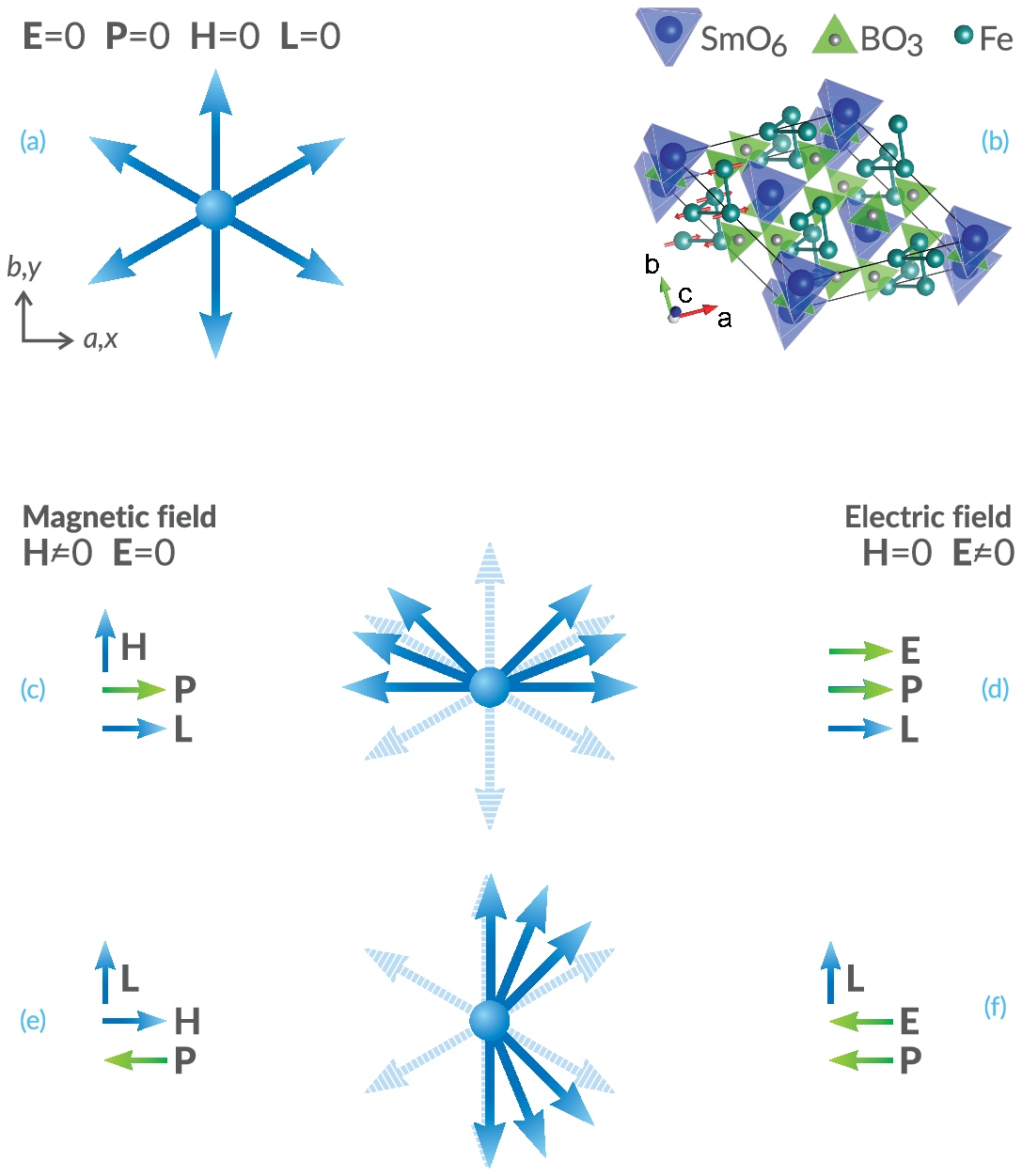}
\end{center}
\caption{\emph{Electric and magnetic ordering in rare-earth iron borates.} (a) Homogeneous distribution of Fe spins (blue arrows) in the $ab$-plane and in the absence of static magnetic ($\mathbf{H}$) and electric ($\mathbf{E}$) fields. Different arrows refer to different domains in the sample. Both, antiferromagnetic vector $\mathbf{L}$ and static polarization $\mathbf{P}$ equal zero in this case. (c,d) Either magnetic field $\mathbf{H}\|b$ (c) or electric field $\mathbf{E}\upuparrows a$ (d) induce  $\mathbf{P}\upuparrows a$ and $\mathbf{L}\|a$. (e,f) Rotation of the magnetic field to $\mathbf{H}\|a$ (e) or inversion of the electric field to $\mathbf{E} \uparrow \downarrow a$ (f) leads to the inversion of the static polarization and the rotation of the antiferromagnetic vector. Panel (b) shows the crystal structure of \Sm.} \label{fig1}
\end{figure}

In zero-field, the magnetic moments of different domains or regions are distributed approximately homogeneously, as illustrated schematically in Fig.~\ref{fig1}(a), thus averaging the electric polarization to zero. The magnetic fields as weak as 0.5~T are enough to break the homogeneous distribution, which leads to a nonzero electric polarization~\cite{zvezdin_jetpl_2006, kadomtseva_ltp_2010, popov_jetp_2010, mukhin_jetpl_2011} according to  Eq.~(\ref{eqPL}) and Figs.~\ref{fig1}(c,e). This effect is quadratic in small magnetic fields and may be described as a first order magnetoelectric effect. Due to the symmetry of the magnetoelectric coupling~\cite{dell_book}, the opposite effect must be possible as well: the magnetization must be sensitive to an external electric field. Indeed, such sensitivity has been recently demonstrated~\cite{freidman_ftt_2015, partzsch_prb_2016} in static experiments for \Sm~ and for \Nd.

Multiferroic iron borates present a rich collection of excitations in the terahertz range~\cite{kuzmenko_jetpl_2011, kuzmenko_jetp_2011, kuzmenko_prb_2014, kuzmenko_prb_2015}.  According the optical experiments~\cite{popova_prb_2007, chukalina_pla_2010}, in the iron borates the splitting of the ground rare-earth doublets are close to the magnon frequencies of the magnetic Fe-subsystem. Therefore, not only the static properties of the iron borates are strongly influenced by the rare-earth~\cite{zvezdin_jetpl_2006, popov_jetp_2010, mukhin_jetpl_2011, popov_prb_2013}, but also the magnetic modes in these systems are strongly coupled. The last effect is seen experimentally as, e.g., a redistribution of the mode intensities and shifts of the resonance frequencies~\cite{kuzmenko_jetpl_2011, kuzmenko_jetp_2011}.

Our experiments revealed that only coupled Fe-rare-earth modes show measurable sensitivity to static electric fields. The strongest effect has been detected for the Sm-Fe mode around \cm{10}. In case of \Sm~ other modes~\cite{kuzmenko_jetpl_2011} may be also expected to reveal voltage sensitivity. For the low-frequency electromagnon~\cite{kuzmenko_prb_2014, kuzmenko_prb_2015} strong static magnetic field must be applied to raise the resonance frequency up to the millimeter frequency range. Magnetic field thus would align the Fe moments (see Fig.~\ref{fig1}) suppressing the voltage effect. The mode around \cm{14}~ is too weak to reveal observable modulation. The high-frequency mode of Sm around~\cm{16} has wrong excitation conditions ($h \| c$-axis) for which it is not sensitive to a rotation of spins in the $ab$-plane. In case of \Nd~ for the Fe mode around \cm{4} no effect could be observed due to the weakness of this excitation.

Terahertz transmission experiments were carried out using quasi-optical terahertz spectroscopy \cite{volkov_infrared_1985,kuzmenko_prb_2016, suppl}. Single crystals of \Sm~ and \Nd~ with typical dimensions of $\sim 1$
cm, were grown by crystallization from the melt on seed
as described in Ref.~[\onlinecite{gudim_crr_2008}].

\begin{figure}[tbp]
\begin{center}
\includegraphics[width=0.99\linewidth, clip]{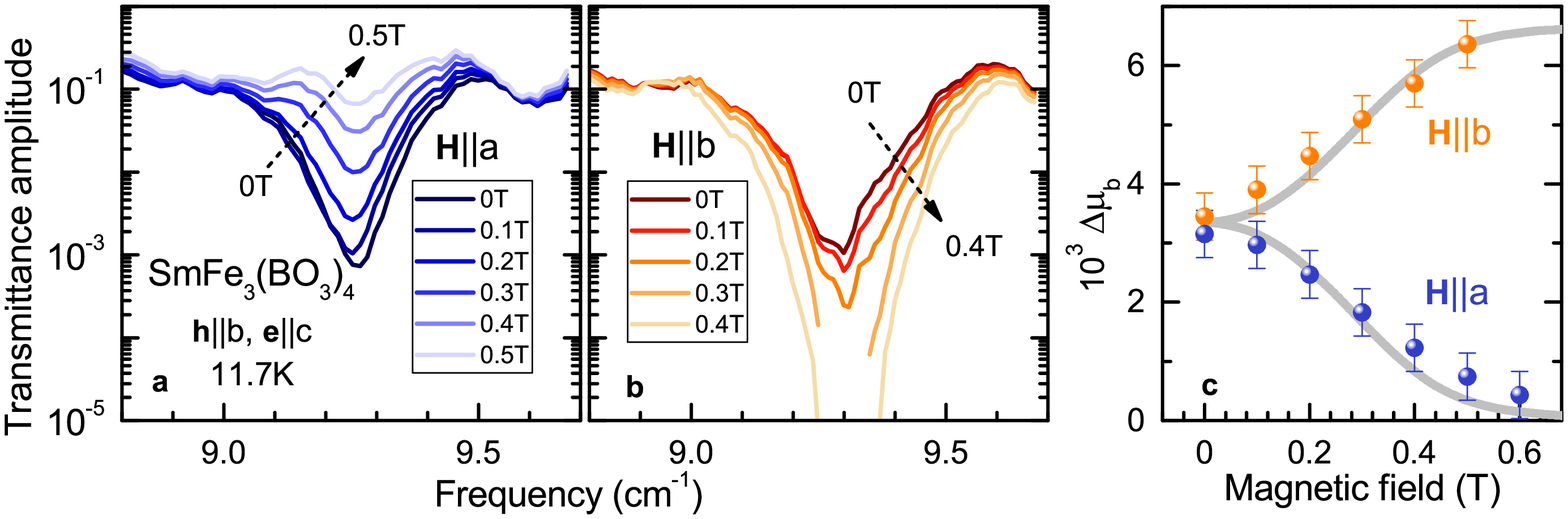}
\end{center}
\caption{\emph{Manipulation of magnon excitation by magnetic field} (a) Suppression of the magnon at \cm{9}~ in \Sm~ by external magnetic fields along the $a$-axis. (b) Increasing of the mode intensity in magnetic fields $\mathbf{H}\|b$-axis. (c) Magnetic field-dependence of the mode intensity from fitting the spectra in (a) and (b) by a Lorentzian function (blue and orange spheres, respectively) and from model calculation based on Eq.~(\ref{model}) (solid gray lines). } \label{fig_mag}
\end{figure}

In \Sm, the coupled Fe-Sm antiferromagnetic mode around \cm{10}~ is of purely magnetic character and it may be excited by an $ac$ magnetic field perpendicular to the antiferromagnetic $\mathbf{L}$-vector~\cite{kuzmenko_jetpl_2011, kuzmenko_jetp_2011}. In the notations of Fig.~\ref{fig1}(a) and without external fields the local magnetic moments are homogeneously distributed in the $ab$-plane. This means that an average of 50 \% of magnetic moments is excited for any orientation of the $ac$ magnetic field in the $ab$-plane. The situation changes drastically if external magnetic or electric fields within the $ab$-plane are present. As demonstrated in Figs.~\ref{fig1}(c-f), external fields destroy the homogeneous distribution of the magnetic moments in the $ab$-plane. In the experiment, this breaks the balance between the excitation conditions with $\mathbf{h}\| a$ and $\mathbf{h}\| b$, respectively, thus shifting the mode intensity to one or the other direction ($\mathbf{h}$ and $\mathbf{e}$ refer to the oscillating magnetic and electric field of light, respectively).

The control of the observed mode intensity by an external magnetic fields is shown in Fig.~\ref{fig_mag} where panels (a,b) demonstrate that the mode strength may be either suppressed or increased depending on the direction of the external magnetic field. As the fields above 0.5~Tesla are sufficient to orient the magnetic moments fully, the intensity of the mode is either saturated at the doubled value compared to $\mathbf{H}=0$ case (Fig.\ref{fig_mag}(b)) or it is suppressed to zero (Fig.\ref{fig_mag}(a)). As follows from the scheme of Fig.~\ref{fig1} and as demonstrated experimentally~\cite{popov_jetp_2010, kadomtseva_ltp_2010}, in both cases either positive or negative static electric polarization is observed along the crystallographic $a$-axis.

\begin{figure}[tbp!]
\begin{center}
\includegraphics[width=0.99\linewidth, clip]{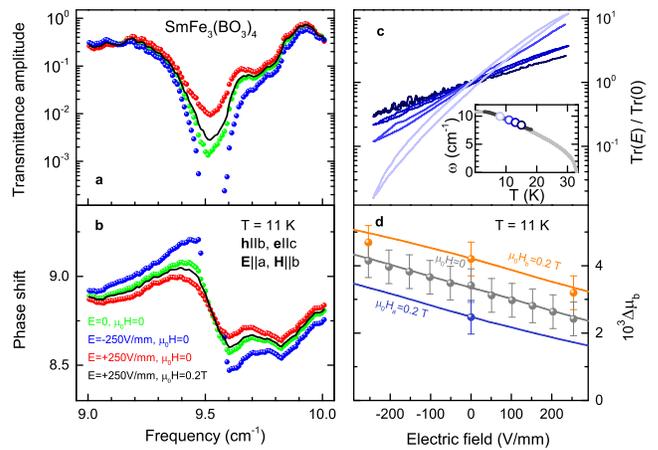}
\end{center}
\caption{\emph{Manipulation of the magnon in \Sm~ by static electric field.} (a) Transmittance amplitude and (b) phase shift spectra in electric field. Symbols: green - initial state, blue - negative electric field, and red - positive electric field. The black line demonstrates that an external magnetic field can approximately compensate the effect of the electric field. (c) Direct modulation of the transmittance amplitude signal by a static electric field at different frequencies. The inset shows the temperature dependence of the resonance frequency as observed (dark gray line) and calculated (light gray line) in Ref.~[\onlinecite{kuzmenko_jetpl_2011}]. The temperatures and frequencies of the transmittance amplitude measurements are marked in the inset by circles. (d) Changes in magnon contribution in the electric field. Symbols are experimental results while the solid lines come from model calculation based on Eq.~\ref{model}. The orange and blue symbols correspond to a simultaneous application of electric and magnetic fields.} \label{fig_el}
\end{figure}

The coupling of electric polarization with an external magnetic field in multiferroic iron borates provides the main idea how to control the magnetic excitations by an electric voltage. By different configurations shown in Fig.~\ref{fig1}, the application of a static voltage along the $a$-axis would favor one of the two possible orientations of the electric polarization. Simultaneously with the static magnetic configurations the excitation conditions for the selected coupled Sm-Fe mode are changed which may be employed for electric field control of the dynamic magnetic properties.


The basic results on electric field control of the magnetic excitation in \Sm~ are shown in Fig.~\ref{fig_el}.  In addition to the magnetic field dependence presented in Fig.~\ref{fig_mag}, close to the resonance position of about \cm{9.5} we observe strong dependence both of the transmittance amplitude and of the phase shift in the electric fields of $\sim 2.5$~kV/cm. Particularly in the case of transmittance amplitude we observe more than one order of magnitude changes in the terahertz signal as influenced by the electric field.

\begin{figure}[tbp!]
\begin{center}
\includegraphics[width=0.99\linewidth, clip]{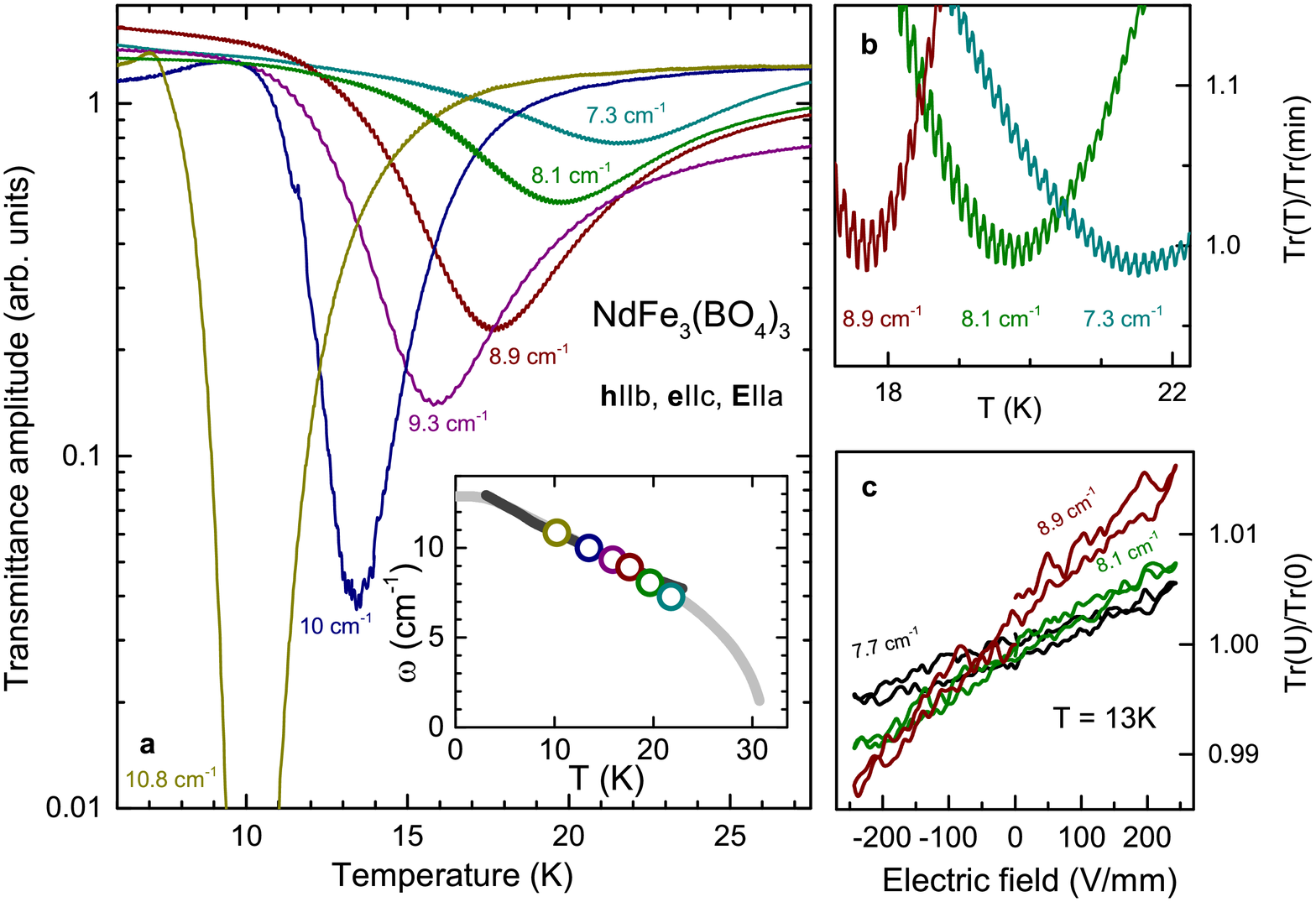}
\end{center}
\caption{\emph{Electric field-effect in \Nd.} (a) Temperature dependence of the relative transmittance amplitude signal for different frequencies. Weak saw-tooth modulation of the curves is due to $\pm 500$~V sweeping of electric voltage during the cooling process. The inset shows the temperature dependence of the resonance frequency as observed (dark gray line) and calculated (light gray line) in Ref.~[\onlinecite{kuzmenko_jetpl_2011}]. The temperatures and frequencies of the resonances seen in (a) are marked in the inset by circles. (b) An example of a detailed view of the data in (a). Here the spectra are normalized by the minimal transmitted intensity. (c) Direct modulation of the terahertz transmittance amplitude at selected frequencies in \Nd.} \label{fig_Nd}
\end{figure}

In spite of the large spectral changes close to the resonance frequency, far from the resonance we observe no measurable changes in the signal. This is due to the fact, that the contribution of the present magnetic mode, shown in Fig.~\ref{fig_el}(d), is small as compared to unity, the relative magnetic permeability of vacuum. In the scale of Fig.~\ref{fig_el}(b), the changes of the optical length of the sample far below the resonance can be estimated as $\Delta l \sim 1 \cdot 10^{-2}$ mm, which is below the sensitivity of the setup.
On the other hand, the $\Delta \mu=3.4 \cdot 10^{-3}$ contribution of the resonance under study and the increase of $\Delta \mu$ in higher fields agrees with the behaviour of the static susceptibility~\cite{demidov_jetp_2013}.
The electric field modulation of magnetic susceptibility in the dynamic regime $d(\Delta \mu)/dE \approx 4\cdot10^{-7}$cm/V is directly connected to the static magnetic susceptibility via $d(\Delta \mu)/dE = d \chi_y/dE_x$. The static values in \Sm~ were recently measured~\cite{freidman_ftt_2015} giving $d \chi_y/dE_x = 2.5\cdot 10^{-8}$cm/V, which is about an order of magnitude lower in value. The simplest explanation of this deviation would be to attribute the static result to the sample twinning, which leads to the suppression of the magnetoelectric signal. However, other mechanisms, such as domain wall motion, cannot be excluded.


The influence of electric and magnetic fields on the magnetic mode contribution $\Delta \mu_y \sim \Delta\mu_0 \left\langle L_x \right\rangle^2$  is determined by the square of the $x$-component of the antiferromagnetic moment $\left\langle L_x \right\rangle^2$, averaged over the sample. To clarify this effect in more detail, we analyzed the actual part of the Landau free energy corresponding to the vector $\mathbf{L}=\left(\cos\varphi,\sin\varphi,0\right)$  in the $xy$-plane\cite{mukhin_jetpl_2011} (vector components are considered in the $x,y,z$ Cartesian basis):
\begin{eqnarray}
\Phi\left(\varphi,E,H\right)=\frac{1}{6}K_6 \cos 6\varphi - \frac{1}{2}K_{1u} \cos 2\varphi - \frac{1}{2}K_{2u} \sin 2\varphi\nonumber \\
 - \frac{1}{2}\chi_\perp H^2 \sin^2\left(\varphi-\varphi_H\right)-P_\perp\left(E_x \cos 2\varphi - E_y\sin 2\varphi \right).
 \label{model}
\end{eqnarray}
Here the first term represents the crystallographic hexagonal anisotropy energy, while the second and third terms stand for the magnetoelastic anisotropy $K_{1u}\sim\sigma_{xx}-\sigma_{yy}$, $K_{2u}\sim\sigma_{xy}$, which are induced by the internal elastic stress of compression/elongation ($\sigma_{xx}-\sigma_{yy}$) and $\sigma_{xy}$ in the $ab$-plane of a real crystal. The fourth term determines the Zeeman energy due to the canting of the antiferromagnetic structure in the magnetic field $\mathbf{H}=H\left(\cos\varphi_H,\sin\varphi_H,0\right)$ and results in $\varphi = \varphi_H\pm \pi/2$ when the magnitude of magnetic field dominates the $ab$-plane anisotropy and the effect of electric field. The last term of Eq.~(\ref{model}) accounts for the magnetoelectric coupling, i.e., the interaction of the spontaneous polarization $\mathbf{P}=\left(P_\perp\cos2\varphi,-P_\perp\sin2\varphi,0\right)$ with  external electric fields. The amplitude of the spontaneous polarization is determined by the magnetoelectric coupling constant and the electric susceptibility as described in Ref.~[\onlinecite{mukhin_jetpl_2011,suppl}]. By minimizing the free energy and taking into account that the crystallographic hexagonal anisotropy is small\cite{mukhin_jetpl_2011} compared to other contributions in Eq.~(\ref{model}), one can find the local orientation of the vector $L$ in the $ab$-plane as a function of electric and magnetic fields:
\begin{equation}
\tan 2\varphi = \frac{ 2 K_{2u}-\chi_\perp H^2 \sin\left(2\varphi_H\right)-4 P_\perp E_y}{2 K_{1u}-\chi_\perp H^2 \cos\left(2\varphi_H\right)+4 P_\perp E_x}.
\label{varphi}
\end{equation}
Assuming random distribution of the magnetoelastic anisotropies $K_{1u}$ and $K_{2u}$ obeying a two-dimensional Gaussian curve we have simulated the behavior of $\Delta \mu_y$ in magnetic and electric fields. These results are shown in Figs.~\ref{fig_mag}(c) and \ref{fig_el}(d) which demonstrate a good description of the experiment. The main parameters of the model were taken from Ref.~[\onlinecite{mukhin_jetpl_2011}] (mean square deviation of the anisotropy $\Delta K_{1u}=\Delta K_{2u}\approx 5.5\times 10^3 \textrm{ erg/cm}^3$ and the transverse magnetic susceptibility $\chi_\perp = 1.2\times 10^{-4} \textrm{ cm}^3\textrm{/g}$), while the maximal value of the spontaneous electric polarization was taken as $P_\perp\approx -240$~$\mu\textrm{C/m}^2$. This value is slightly lower than that observed in Refs.~[\onlinecite{popov_jetp_2010,mukhin_jetpl_2011}] likely due to a larger amount of crystallographic inversion twins in the enantiomorph crystal.
Remarkably, according to Eq.~(\ref{varphi}), the simultaneous application of both $\mathbf{E}$ and $\mathbf{H}$ could lead to a compensation of their action as a result of an interrelation between them. For example, for $\mathbf{E}\parallel a$ and $\mathbf{H}\parallel b$ the compensation effect occurs according to $\chi_\perp H^2 + 4P_\perp E_x = 0$, which is in a good agreement with our measurements for $E=+250\textrm{ V/mm}$ and $\mu_0 H_b = 0.2\textrm{ T}$ (Figs.~\ref{fig_el}(a,b,d)).

Figure~\ref{fig_Nd} shows typical results of electric field experiments in \Nd. Panel (a) demonstrates the temperature dependence of the transmittance amplitude at selected frequencies. Characteristic minima in these data correspond to a crossing of the temperature-dependent resonance frequency of the mode and the frequency of the experiment, as shown in the inset. These measurements were obtained with cooling at 1~K/min and the simultaneous sweeping of the gate voltage between -500~V and +500~V at a rate of $\sim 0.1$ Hz. The characteristic saw-tooth profile of these curves demonstrate the nonzero effect of the electric field on this magnetic mode in \Nd.
 From the slopes shown in Fig.~\ref{fig_Nd}(c) the field-dependent susceptibility may be estimated as $d \chi_y/dE_x = 2.4\cdot 10^{-8}$cm/V, which is an order of magnitude smaller than the same values from \Sm. This difference is due to a small value of the spontaneous electric polarization and to larger threshold magnetic field to suppress the spiral magnetic structure in \Nd\ ($\sim 1$~T compared to $\sim 0.3$~T for \Sm)\cite{kadomtseva_ltp_2006, popov_jetp_2010}.

The reaction time of the present experimental setup can be estimated as  $\sim 45$~ms. Within this time scale an instantaneous response of the the magnetic system to the changes of electric field have been observed. Based on the $ac$ results given in Ref.~[\onlinecite{freidman_ftt_2015}] the switching times of at most 1~ms may be expected. As mentioned in the introduction, in case of domain wall motion the switching time of the devices are limited by tenths of nanoseconds. In magnetoelectric ferroborates the process includes both, rotation of the magnetic moments and switching of the electric polarization.
The characteristic time scale for the magnetic part is determined by the in-plane antiferromagnetic resonance frequency ($\sim 5$\,GHz at $H=0$)\cite{kuzmenko_prb_2014}, which will probably determine the switching rate. Finally, for short pulses, electric and magnetic fields are present simultaneously. This mixing may influence the switching on the short time scales.

In conclusion, magnetic modes in multiferroic ferroborates are shown to be sensitive to both, external magnetic field and static voltage.  Nearly 100\,\% modulation of the terahertz radiation in an external electric field is demonstrated for \Sm. The experimental results can be well explained using a theoretical model which includes the magnetoelectric coupling in multiferroic borates.  High sensitivity to electric voltage is due to a strong effect of both magnetic and electric fields on the spin orientation in an easy plane antiferromagnetic structure and significant coupling of the rare-earth and the iron magnetic subsystems.


\subsection*{Acknowledgements}
This work was supported by the Russian Science Foundation
(16-12-10531: AMK, VYuI and AAM), by the Russian Foundation for Basic
Research 17-52-45091 IND-a : IAG and LNB), and by
the Austrian Science Funds (W1243, I 2816-N27, I 1648-N27).

\subsection*{Supplementary Information}

\subsubsection*{Multiferroic borates}

Rare-earth iron-borates represent one exotic class of multiferroics~\cite{vasiliev_ltp_2006, kadomtseva_ltp_2010, kadomtseva_jetp_2012}. At high temperatures, all rare-earth borates reveal a non-centrosymmetric trigonal structure belonging to the space group R32~\cite{campa_chemm_1997, hinatsu_jssch_2002} which persist down to lowest temperatures for compounds with large ionic radius of the rare-earth (La-Sm). In iron borates with smaller ionic radius (Eu-Er, Y) a phase transition to the structure within P$3_121$ space group takes place for lower temperatures~\cite{fausti_prb_2006, popova_jre_2009}.

In multiferroic iron borates the coupling of the paramagnetic rare earth (R) and antiferromagnetically ordered Fe-moments arises due to the exchange interaction and it results in an induced antiferromagnetic order in the Sm subsystem. For the easy-plane ground state of Fe-spins in \Sm~ the orientation of the Sm magnetic moments also occurs in the easy ab-plane and the Sm-order takes place at the N\'{e}el temperature of the Fe-subsystem $\sim 30$~K. Strictly speaking, this is the ordering temperature of both, Fe- and Sm- subsystems. However, since the Fe-Fe exchange interaction is much stronger than the Sm-Fe, it is reasonable to consider the Fe ordering as a primary order parameter.

The role of anisotropy of the R-subsystem is very important in determining  the orientation of the iron spins.
In the case of \Sm~  the ground doublet of Sm$^{3+}$ is split by the Sm-Fe exchange interaction thus stabilizing the easy-plane state~\cite{kuzmenko_jetpl_2011}. In case of \Nd~ the collinear easy-plane state below the N\'{e}el temperature is transformed into the spiral easy-plane state around 13-15\,K. The origin of this transition still remains unclear. In several other iron borates like GdFe$_3$(BO$_3$)$_4$ and HoFe$_3$(BO$_3$)$_4$ the spin reorientation from the easy plane to the easy axis state exists due to competitions of the Fe- and R-subsystem anisotropies~\cite{vasiliev_ltp_2006, kadomtseva_ltp_2010, kadomtseva_jetp_2012}.

\subsubsection*{Terahertz spectroscopy}

Terahertz transmission experiments were carried out using quasi-optical terahertz spectroscopy \cite{volkov_infrared_1985}. This technique utilizes linearly polarized monochromatic radiation provided by backward-wave-oscillators. He-cooled bolometers were used as detectors of the radiation. Using wire grid polarizers, the complex transmission
coefficient can be obtained both in parallel and crossed polarizers
geometry. The phase information was obtained by comparing the mirror positions necessary to reach an interference minimum between the two arms of our Mach-Zender interferometer for the sample and for the reference aperture. Static magnetic fields, up to $\pm7$~Tesla, have been
applied to the sample using a split-coil superconducting magnet.
Frequency dependent transmission spectra were analyzed using the Fresnel optical formulas for the transmittance of a plane-parallel sample~\cite{kuzmenko_prb_2016} assuming a Lorentzian form of the magnetic excitations $\mu(\omega)=1+ \Delta \mu \omega_0^2/(\omega_0^2-\omega^2-i\omega g)$.  Here $\Delta \mu$ is the mode contribution, $\omega_0$ is the resonance frequency, $g$ is the mode width, and $\omega$ is the angular frequency of the experiment.

\subsubsection*{Magnetic interactions}
The magnetic part of the thermodynamic potential in Eq.~(\ref{model}) of the main text is derived from Eq.~(2) of Ref.~[\onlinecite{mukhin_jetpl_2011}] for the special case of magnetic moments restricted to the $xy$ magnetic easy-plane. Considering the antiferromagnetic vector as $\mathbf{L}=\left(\cos\varphi,\sin\varphi,0\right)$, and the external magnetic field as $\mathbf{H}=H\left(\cos\varphi_H,\sin\varphi_H,0\right)$, both lying in the $xy$-plane, Eq.~(2) of Ref.~[\onlinecite{mukhin_jetpl_2011}] takes the form
\begin{eqnarray}
\Phi_m\left(\varphi,H\right) &=& \frac{1}{6}K_6 \cos 6\varphi - \frac{1}{2}K_{1u} \cos 2\varphi  - \frac{1}{2}K_{2u} \sin 2\varphi \nonumber \\ &-& \frac{1}{2}\chi_\perp H^2 \sin^2\left(\varphi-\varphi_H\right),
 \label{Phi_m}
\end{eqnarray}
which is identical to the magnetic part of Eq.~(\ref{model}) of the main text.

\subsubsection*{Magnetoelectric coupling}

The magnetoelectric term of Eq.~(\ref{model}) in the main text is a simplified form of the more general expressions of Ref.~[\onlinecite{mukhin_jetpl_2011}]. In Ref.~[\onlinecite{mukhin_jetpl_2011}] the Eqs.~(3) and (4) give the magnetoelectric and electric part of the thermodynamic potential, respectively, as
\begin{eqnarray}
\Phi_{me}\left(\varphi,P\right)+\Phi_{e}\left(P,E\right)&=& -c_2\left(P_x \cos 2\varphi - P_y \sin 2\varphi \right) \nonumber \\ &+& \frac{P_x^2+P_y^2}{2\chi_\perp^e} -\mathbf{P}\mathbf{E}.
 \label{Phi_ME}
\end{eqnarray}
The first term of the right-hand side represents the magnetoelectric coupling between the magnetic order, i.e. $\varphi$, and the $\mathbf{P}$ electric polarization. Here $c_2$ is the magnetoelectric coupling constant, while in the electric term $\chi_\perp^e$ denotes the electric susceptibility of the crystal in the $xy$-plane and $\mathbf{E}$ is the external electric field. For a given magnetic order in the $xy$-plane, i.e. for a given $\varphi$, the polarization minimizing the thermodynamic potential reads as~\cite{mukhin_jetpl_2011}
\begin{eqnarray}
P_x&=&P_\perp\cos 2\varphi + \chi_\perp^e E_x \\
P_y&=&-P_\perp\sin 2\varphi + \chi_\perp^e E_y,
\label{P}	
\end{eqnarray}
where the amplitude of the spontaneous polarization is $P_\perp=c_2\chi_\perp^e$. Omitting the terms from Eq.~(\ref{Phi_ME}) which are independent of the spin configuration, i.e. independent of $\varphi$, we arrive at the $-P_\perp\left(E_x \cos 2\varphi - E_y\sin 2\varphi \right)$ magnetoelectric term of Eq.~(\ref{model}).

\bibliography{literature}

\begin{thebibliography}{72}
\expandafter\ifx\csname natexlab\endcsname\relax\def\natexlab#1{#1}\fi
\expandafter\ifx\csname bibnamefont\endcsname\relax
  \def\bibnamefont#1{#1}\fi
\expandafter\ifx\csname bibfnamefont\endcsname\relax
  \def\bibfnamefont#1{#1}\fi
\expandafter\ifx\csname citenamefont\endcsname\relax
  \def\citenamefont#1{#1}\fi
\expandafter\ifx\csname url\endcsname\relax
  \def\url#1{\texttt{#1}}\fi
\expandafter\ifx\csname urlprefix\endcsname\relax\def\urlprefix{URL }\fi
\providecommand{\bibinfo}[2]{#2}
\providecommand{\eprint}[2][]{\url{#2}}

\bibitem[{\citenamefont{Vaz}(2012)}]{vaz_jpcm_2012}
\bibinfo{author}{\bibfnamefont{C.~A.~F.} \bibnamefont{Vaz}},
  \bibinfo{journal}{J. Phys.: Condens. Matter} \textbf{\bibinfo{volume}{24}},
  \bibinfo{pages}{333201} (\bibinfo{year}{2012}),
  \urlprefix\url{http://stacks.iop.org/0953-8984/24/i=33/a=333201}.

\bibitem[{\citenamefont{Matsukura et~al.}(2015)\citenamefont{Matsukura, Tokura,
  and Ohno}}]{matsukura_natnano_2015}
\bibinfo{author}{\bibfnamefont{F.}~\bibnamefont{Matsukura}},
  \bibinfo{author}{\bibfnamefont{Y.}~\bibnamefont{Tokura}}, \bibnamefont{and}
  \bibinfo{author}{\bibfnamefont{H.}~\bibnamefont{Ohno}},
  \bibinfo{journal}{Nat. Nanotech.} \textbf{\bibinfo{volume}{10}},
  \bibinfo{pages}{209} (\bibinfo{year}{2015}),
  \urlprefix\url{http://www.nature.com/nnano/journal/v10/n3/abs/nnano.2015.22.html}.

\bibitem[{\citenamefont{Smolenskii and Chupis}(1982)}]{smolenskii_ufn_1982}
\bibinfo{author}{\bibfnamefont{G.~A.} \bibnamefont{Smolenskii}}
  \bibnamefont{and} \bibinfo{author}{\bibfnamefont{I.~E.}
  \bibnamefont{Chupis}}, \bibinfo{journal}{Sov. Phys. Usp.}
  \textbf{\bibinfo{volume}{25}}, \bibinfo{pages}{475} (\bibinfo{year}{1982}),
  \urlprefix\url{http://stacks.iop.org/0038-5670/25/i=7/a=R02}.

\bibitem[{\citenamefont{Fiebig}(2005)}]{fiebig_jpd_2005}
\bibinfo{author}{\bibfnamefont{M.}~\bibnamefont{Fiebig}}, \bibinfo{journal}{J.
  Phys. D: Appl. Phys.} \textbf{\bibinfo{volume}{38}}, \bibinfo{pages}{R123}
  (\bibinfo{year}{2005}),
  \urlprefix\url{http://stacks.iop.org/0022-3727/38/R123}.

\bibitem[{\citenamefont{Dong et~al.}(2015)\citenamefont{Dong, Liu, Cheong, and
  Ren}}]{dong_advph_2015}
\bibinfo{author}{\bibfnamefont{S.}~\bibnamefont{Dong}},
  \bibinfo{author}{\bibfnamefont{J.-M.} \bibnamefont{Liu}},
  \bibinfo{author}{\bibfnamefont{S.-W.} \bibnamefont{Cheong}},
  \bibnamefont{and} \bibinfo{author}{\bibfnamefont{Z.}~\bibnamefont{Ren}},
  \bibinfo{journal}{Adv. Phys.} \textbf{\bibinfo{volume}{64}},
  \bibinfo{pages}{519} (\bibinfo{year}{2015}),
  \urlprefix\url{http://dx.doi.org/10.1080/00018732.2015.1114338}.

\bibitem[{\citenamefont{Tokura et~al.}(2014)\citenamefont{Tokura, Seki, and
  Nagaosa}}]{tokura_rpp_2014}
\bibinfo{author}{\bibfnamefont{Y.}~\bibnamefont{Tokura}},
  \bibinfo{author}{\bibfnamefont{S.}~\bibnamefont{Seki}}, \bibnamefont{and}
  \bibinfo{author}{\bibfnamefont{N.}~\bibnamefont{Nagaosa}},
  \bibinfo{journal}{Rep. Prog. Phys.} \textbf{\bibinfo{volume}{77}},
  \bibinfo{pages}{076501} (\bibinfo{year}{2014}),
  \urlprefix\url{http://stacks.iop.org/0034-4885/77/i=7/a=076501}.

\bibitem[{\citenamefont{Fiebig et~al.}(2016)\citenamefont{Fiebig, Lottermoser,
  Meier, and Trassin}}]{fiebig_natrev_2016}
\bibinfo{author}{\bibfnamefont{M.}~\bibnamefont{Fiebig}},
  \bibinfo{author}{\bibfnamefont{T.}~\bibnamefont{Lottermoser}},
  \bibinfo{author}{\bibfnamefont{D.}~\bibnamefont{Meier}}, \bibnamefont{and}
  \bibinfo{author}{\bibfnamefont{M.}~\bibnamefont{Trassin}},
  \bibinfo{journal}{Nat. Rev. Mater.} \textbf{\bibinfo{volume}{1}},
  \bibinfo{pages}{16046} (\bibinfo{year}{2016}).

\bibitem[{\citenamefont{Lottermoser et~al.}(2004)\citenamefont{Lottermoser,
  Lonkai, Amann, Hohlwein, Ihringer, and Fiebig}}]{lottermoser_nature_2004}
\bibinfo{author}{\bibfnamefont{T.}~\bibnamefont{Lottermoser}},
  \bibinfo{author}{\bibfnamefont{T.}~\bibnamefont{Lonkai}},
  \bibinfo{author}{\bibfnamefont{U.}~\bibnamefont{Amann}},
  \bibinfo{author}{\bibfnamefont{D.}~\bibnamefont{Hohlwein}},
  \bibinfo{author}{\bibfnamefont{J.}~\bibnamefont{Ihringer}}, \bibnamefont{and}
  \bibinfo{author}{\bibfnamefont{M.}~\bibnamefont{Fiebig}},
  \bibinfo{journal}{Nature (London)} \textbf{\bibinfo{volume}{430}},
  \bibinfo{pages}{541} (\bibinfo{year}{2004}).

\bibitem[{\citenamefont{Chu et~al.}(2008)\citenamefont{Chu, Martin, Holcomb,
  Gajek, Han, He, Balke, Yang, Lee, Hu et~al.}}]{chu_natmat_2008}
\bibinfo{author}{\bibfnamefont{Y.-H.} \bibnamefont{Chu}},
  \bibinfo{author}{\bibfnamefont{L.~W.} \bibnamefont{Martin}},
  \bibinfo{author}{\bibfnamefont{M.~B.} \bibnamefont{Holcomb}},
  \bibinfo{author}{\bibfnamefont{M.}~\bibnamefont{Gajek}},
  \bibinfo{author}{\bibfnamefont{S.-J.} \bibnamefont{Han}},
  \bibinfo{author}{\bibfnamefont{Q.}~\bibnamefont{He}},
  \bibinfo{author}{\bibfnamefont{N.}~\bibnamefont{Balke}},
  \bibinfo{author}{\bibfnamefont{C.-H.} \bibnamefont{Yang}},
  \bibinfo{author}{\bibfnamefont{D.}~\bibnamefont{Lee}},
  \bibinfo{author}{\bibfnamefont{W.}~\bibnamefont{Hu}}, \bibnamefont{et~al.},
  \bibinfo{journal}{Nat. Mater.} \textbf{\bibinfo{volume}{7}},
  \bibinfo{pages}{478} (\bibinfo{year}{2008}).

\bibitem[{\citenamefont{Saito et~al.}(2009)\citenamefont{Saito, Ishikawa,
  Konno, Taniguchi, and Arima}}]{saito_nmat_2009}
\bibinfo{author}{\bibfnamefont{M.}~\bibnamefont{Saito}},
  \bibinfo{author}{\bibfnamefont{K.}~\bibnamefont{Ishikawa}},
  \bibinfo{author}{\bibfnamefont{S.}~\bibnamefont{Konno}},
  \bibinfo{author}{\bibfnamefont{K.}~\bibnamefont{Taniguchi}},
  \bibnamefont{and} \bibinfo{author}{\bibfnamefont{T.}~\bibnamefont{Arima}},
  \bibinfo{journal}{Nat. Mater.} \textbf{\bibinfo{volume}{8}},
  \bibinfo{pages}{634} (\bibinfo{year}{2009}).

\bibitem[{\citenamefont{Tokunaga et~al.}(2009)\citenamefont{Tokunaga, Furukawa,
  Sakai, Taguchi, Arima, and Tokura}}]{tokunaga_natmat_2009}
\bibinfo{author}{\bibfnamefont{Y.}~\bibnamefont{Tokunaga}},
  \bibinfo{author}{\bibfnamefont{N.}~\bibnamefont{Furukawa}},
  \bibinfo{author}{\bibfnamefont{H.}~\bibnamefont{Sakai}},
  \bibinfo{author}{\bibfnamefont{Y.}~\bibnamefont{Taguchi}},
  \bibinfo{author}{\bibfnamefont{T.-h.} \bibnamefont{Arima}}, \bibnamefont{and}
  \bibinfo{author}{\bibfnamefont{Y.}~\bibnamefont{Tokura}},
  \bibinfo{journal}{Nat. Mater.} \textbf{\bibinfo{volume}{8}},
  \bibinfo{pages}{558} (\bibinfo{year}{2009}).

\bibitem[{\citenamefont{Choi et~al.}(2010)\citenamefont{Choi, Zhang, Lee, and
  Cheong}}]{choi_prl_2010}
\bibinfo{author}{\bibfnamefont{Y.~J.} \bibnamefont{Choi}},
  \bibinfo{author}{\bibfnamefont{C.~L.} \bibnamefont{Zhang}},
  \bibinfo{author}{\bibfnamefont{N.}~\bibnamefont{Lee}}, \bibnamefont{and}
  \bibinfo{author}{\bibfnamefont{S.-W.} \bibnamefont{Cheong}},
  \bibinfo{journal}{Phys. Rev. Lett.} \textbf{\bibinfo{volume}{105}},
  \bibinfo{pages}{097201} (\bibinfo{year}{2010}),
  \urlprefix\url{https://link.aps.org/doi/10.1103/PhysRevLett.105.097201}.

\bibitem[{\citenamefont{Oh et~al.}(2014)\citenamefont{Oh, Artyukhin, Yang,
  Zapf, Kim, Vanderbilt, and Cheong}}]{oh_natcomm_2014}
\bibinfo{author}{\bibfnamefont{Y.~S.} \bibnamefont{Oh}},
  \bibinfo{author}{\bibfnamefont{S.}~\bibnamefont{Artyukhin}},
  \bibinfo{author}{\bibfnamefont{J.~J.} \bibnamefont{Yang}},
  \bibinfo{author}{\bibfnamefont{V.}~\bibnamefont{Zapf}},
  \bibinfo{author}{\bibfnamefont{J.~W.} \bibnamefont{Kim}},
  \bibinfo{author}{\bibfnamefont{D.}~\bibnamefont{Vanderbilt}},
  \bibnamefont{and} \bibinfo{author}{\bibfnamefont{S.-W.}
  \bibnamefont{Cheong}}, \bibinfo{journal}{Nat. Comm.}
  \textbf{\bibinfo{volume}{5}}, \bibinfo{pages}{4201} (\bibinfo{year}{2014}).

\bibitem[{\citenamefont{Soda et~al.}(2016)\citenamefont{Soda, Hayashida,
  Roessli, M\aa{}nsson, White, Matsumoto, Shiina, and Masuda}}]{soda_prb_2016}
\bibinfo{author}{\bibfnamefont{M.}~\bibnamefont{Soda}},
  \bibinfo{author}{\bibfnamefont{S.}~\bibnamefont{Hayashida}},
  \bibinfo{author}{\bibfnamefont{B.}~\bibnamefont{Roessli}},
  \bibinfo{author}{\bibfnamefont{M.}~\bibnamefont{M\aa{}nsson}},
  \bibinfo{author}{\bibfnamefont{J.~S.} \bibnamefont{White}},
  \bibinfo{author}{\bibfnamefont{M.}~\bibnamefont{Matsumoto}},
  \bibinfo{author}{\bibfnamefont{R.}~\bibnamefont{Shiina}}, \bibnamefont{and}
  \bibinfo{author}{\bibfnamefont{T.}~\bibnamefont{Masuda}},
  \bibinfo{journal}{Phys. Rev. B} \textbf{\bibinfo{volume}{94}},
  \bibinfo{pages}{094418} (\bibinfo{year}{2016}),
  \urlprefix\url{http://link.aps.org/doi/10.1103/PhysRevB.94.094418}.

\bibitem[{\citenamefont{Yamasaki et~al.}(2007)\citenamefont{Yamasaki, Sagayama,
  Goto, Matsuura, Hirota, Arima, and Tokura}}]{yamasaki_prl_2007}
\bibinfo{author}{\bibfnamefont{Y.}~\bibnamefont{Yamasaki}},
  \bibinfo{author}{\bibfnamefont{H.}~\bibnamefont{Sagayama}},
  \bibinfo{author}{\bibfnamefont{T.}~\bibnamefont{Goto}},
  \bibinfo{author}{\bibfnamefont{M.}~\bibnamefont{Matsuura}},
  \bibinfo{author}{\bibfnamefont{K.}~\bibnamefont{Hirota}},
  \bibinfo{author}{\bibfnamefont{T.}~\bibnamefont{Arima}}, \bibnamefont{and}
  \bibinfo{author}{\bibfnamefont{Y.}~\bibnamefont{Tokura}},
  \bibinfo{journal}{Phys. Rev. Lett.} \textbf{\bibinfo{volume}{98}},
  \bibinfo{pages}{147204} (\bibinfo{year}{2007}),
  \urlprefix\url{https://link.aps.org/doi/10.1103/PhysRevLett.98.147204}.

\bibitem[{\citenamefont{Murakawa et~al.}(2009)\citenamefont{Murakawa, Onose,
  and Tokura}}]{murakawa_prl_2009}
\bibinfo{author}{\bibfnamefont{H.}~\bibnamefont{Murakawa}},
  \bibinfo{author}{\bibfnamefont{Y.}~\bibnamefont{Onose}}, \bibnamefont{and}
  \bibinfo{author}{\bibfnamefont{Y.}~\bibnamefont{Tokura}},
  \bibinfo{journal}{Phys. Rev. Lett.} \textbf{\bibinfo{volume}{103}},
  \bibinfo{pages}{147201} (\bibinfo{year}{2009}),
  \urlprefix\url{https://link.aps.org/doi/10.1103/PhysRevLett.103.147201}.

\bibitem[{\citenamefont{Finger et~al.}(2010)\citenamefont{Finger, Senff,
  Schmalzl, Schmidt, Regnault, Becker, Bohat\`{y}, and
  Braden}}]{finger_prb_2010}
\bibinfo{author}{\bibfnamefont{T.}~\bibnamefont{Finger}},
  \bibinfo{author}{\bibfnamefont{D.}~\bibnamefont{Senff}},
  \bibinfo{author}{\bibfnamefont{K.}~\bibnamefont{Schmalzl}},
  \bibinfo{author}{\bibfnamefont{W.}~\bibnamefont{Schmidt}},
  \bibinfo{author}{\bibfnamefont{L.~P.} \bibnamefont{Regnault}},
  \bibinfo{author}{\bibfnamefont{P.}~\bibnamefont{Becker}},
  \bibinfo{author}{\bibfnamefont{L.}~\bibnamefont{Bohat\`{y}}},
  \bibnamefont{and} \bibinfo{author}{\bibfnamefont{M.}~\bibnamefont{Braden}},
  \bibinfo{journal}{Phys. Rev. B} \textbf{\bibinfo{volume}{81}},
  \bibinfo{pages}{054430} (\bibinfo{year}{2010}),
  \urlprefix\url{http://link.aps.org/doi/10.1103/PhysRevB.81.054430}.

\bibitem[{\citenamefont{Babkevich et~al.}(2012)\citenamefont{Babkevich, Poole,
  Johnson, Roessli, Prabhakaran, and Boothroyd}}]{babkevich_prb_2012}
\bibinfo{author}{\bibfnamefont{P.}~\bibnamefont{Babkevich}},
  \bibinfo{author}{\bibfnamefont{A.}~\bibnamefont{Poole}},
  \bibinfo{author}{\bibfnamefont{R.~D.} \bibnamefont{Johnson}},
  \bibinfo{author}{\bibfnamefont{B.}~\bibnamefont{Roessli}},
  \bibinfo{author}{\bibfnamefont{D.}~\bibnamefont{Prabhakaran}},
  \bibnamefont{and} \bibinfo{author}{\bibfnamefont{A.~T.}
  \bibnamefont{Boothroyd}}, \bibinfo{journal}{Phys. Rev. B}
  \textbf{\bibinfo{volume}{85}}, \bibinfo{pages}{134428}
  (\bibinfo{year}{2012}),
  \urlprefix\url{https://link.aps.org/doi/10.1103/PhysRevB.85.134428}.

\bibitem[{\citenamefont{Zhao et~al.}(2006)\citenamefont{Zhao, Scholl,
  Zavaliche, Lee, Barry, Doran, Cruz, Chu, Ederer, Spaldin
  et~al.}}]{zhao_nmat_2006}
\bibinfo{author}{\bibfnamefont{T.}~\bibnamefont{Zhao}},
  \bibinfo{author}{\bibfnamefont{A.}~\bibnamefont{Scholl}},
  \bibinfo{author}{\bibfnamefont{F.}~\bibnamefont{Zavaliche}},
  \bibinfo{author}{\bibfnamefont{K.}~\bibnamefont{Lee}},
  \bibinfo{author}{\bibfnamefont{M.}~\bibnamefont{Barry}},
  \bibinfo{author}{\bibfnamefont{A.}~\bibnamefont{Doran}},
  \bibinfo{author}{\bibfnamefont{M.~P.} \bibnamefont{Cruz}},
  \bibinfo{author}{\bibfnamefont{Y.~H.} \bibnamefont{Chu}},
  \bibinfo{author}{\bibfnamefont{C.}~\bibnamefont{Ederer}},
  \bibinfo{author}{\bibfnamefont{N.~A.} \bibnamefont{Spaldin}},
  \bibnamefont{et~al.}, \bibinfo{journal}{Nat. Mater.}
  \textbf{\bibinfo{volume}{5}}, \bibinfo{pages}{823} (\bibinfo{year}{2006}).

\bibitem[{\citenamefont{Evans et~al.}(2013)\citenamefont{Evans, Schilling,
  Kumar, Sanchez, Ortega, Arredondo, Katiyar, Gregg, and
  Scott}}]{evans_natcomm_2013}
\bibinfo{author}{\bibfnamefont{D.~M.} \bibnamefont{Evans}},
  \bibinfo{author}{\bibfnamefont{A.}~\bibnamefont{Schilling}},
  \bibinfo{author}{\bibfnamefont{A.}~\bibnamefont{Kumar}},
  \bibinfo{author}{\bibfnamefont{D.}~\bibnamefont{Sanchez}},
  \bibinfo{author}{\bibfnamefont{N.}~\bibnamefont{Ortega}},
  \bibinfo{author}{\bibfnamefont{M.}~\bibnamefont{Arredondo}},
  \bibinfo{author}{\bibfnamefont{R.~S.} \bibnamefont{Katiyar}},
  \bibinfo{author}{\bibfnamefont{J.~M.} \bibnamefont{Gregg}}, \bibnamefont{and}
  \bibinfo{author}{\bibfnamefont{J.~F.} \bibnamefont{Scott}},
  \bibinfo{journal}{Nat. Comm.} \textbf{\bibinfo{volume}{4}},
  \bibinfo{pages}{1534} (\bibinfo{year}{2013}).

\bibitem[{\citenamefont{Wang et~al.}(2009)\citenamefont{Wang, Liu, and
  Ren}}]{wang_advph_2009}
\bibinfo{author}{\bibfnamefont{K.~F.} \bibnamefont{Wang}},
  \bibinfo{author}{\bibfnamefont{J.-M.} \bibnamefont{Liu}}, \bibnamefont{and}
  \bibinfo{author}{\bibfnamefont{Z.~F.} \bibnamefont{Ren}},
  \bibinfo{journal}{Adv. Phys.} \textbf{\bibinfo{volume}{58}},
  \bibinfo{pages}{321} (\bibinfo{year}{2009}),
  \urlprefix\url{http://dx.doi.org/10.1080/00018730902920554}.

\bibitem[{\citenamefont{Fatuzzo}(1962)}]{fatuzzo_pr_1962}
\bibinfo{author}{\bibfnamefont{E.}~\bibnamefont{Fatuzzo}},
  \bibinfo{journal}{Phys. Rev.} \textbf{\bibinfo{volume}{127}},
  \bibinfo{pages}{1999} (\bibinfo{year}{1962}),
  \urlprefix\url{https://link.aps.org/doi/10.1103/PhysRev.127.1999}.

\bibitem[{\citenamefont{Tybell et~al.}(2002)\citenamefont{Tybell, Paruch,
  Giamarchi, and Triscone}}]{tybell_prl_2002}
\bibinfo{author}{\bibfnamefont{T.}~\bibnamefont{Tybell}},
  \bibinfo{author}{\bibfnamefont{P.}~\bibnamefont{Paruch}},
  \bibinfo{author}{\bibfnamefont{T.}~\bibnamefont{Giamarchi}},
  \bibnamefont{and} \bibinfo{author}{\bibfnamefont{J.-M.}
  \bibnamefont{Triscone}}, \bibinfo{journal}{Phys. Rev. Lett.}
  \textbf{\bibinfo{volume}{89}}, \bibinfo{pages}{097601}
  (\bibinfo{year}{2002}),
  \urlprefix\url{https://link.aps.org/doi/10.1103/PhysRevLett.89.097601}.

\bibitem[{\citenamefont{Dawber et~al.}(2005)\citenamefont{Dawber, Rabe, and
  Scott}}]{dawber_rmp_2005}
\bibinfo{author}{\bibfnamefont{M.}~\bibnamefont{Dawber}},
  \bibinfo{author}{\bibfnamefont{K.~M.} \bibnamefont{Rabe}}, \bibnamefont{and}
  \bibinfo{author}{\bibfnamefont{J.~F.} \bibnamefont{Scott}},
  \bibinfo{journal}{Rev. Mod. Phys.} \textbf{\bibinfo{volume}{77}},
  \bibinfo{pages}{1083} (\bibinfo{year}{2005}),
  \urlprefix\url{https://link.aps.org/doi/10.1103/RevModPhys.77.1083}.

\bibitem[{\citenamefont{Grigoriev et~al.}(2006)\citenamefont{Grigoriev, Do,
  Kim, Eom, Adams, Dufresne, and Evans}}]{grigoriev_prl_2006}
\bibinfo{author}{\bibfnamefont{A.}~\bibnamefont{Grigoriev}},
  \bibinfo{author}{\bibfnamefont{D.-H.} \bibnamefont{Do}},
  \bibinfo{author}{\bibfnamefont{D.~M.} \bibnamefont{Kim}},
  \bibinfo{author}{\bibfnamefont{C.-B.} \bibnamefont{Eom}},
  \bibinfo{author}{\bibfnamefont{B.}~\bibnamefont{Adams}},
  \bibinfo{author}{\bibfnamefont{E.~M.} \bibnamefont{Dufresne}},
  \bibnamefont{and} \bibinfo{author}{\bibfnamefont{P.~G.} \bibnamefont{Evans}},
  \bibinfo{journal}{Phys. Rev. Lett.} \textbf{\bibinfo{volume}{96}},
  \bibinfo{pages}{187601} (\bibinfo{year}{2006}),
  \urlprefix\url{https://link.aps.org/doi/10.1103/PhysRevLett.96.187601}.

\bibitem[{\citenamefont{Ehara et~al.}(2017)\citenamefont{Ehara, Yasui, Oikawa,
  Shiraishi, Shimizu, Tanaka, Kanenko, Maran, Yamada, Imai
  et~al.}}]{ehara_scirep_2017}
\bibinfo{author}{\bibfnamefont{Y.}~\bibnamefont{Ehara}},
  \bibinfo{author}{\bibfnamefont{S.}~\bibnamefont{Yasui}},
  \bibinfo{author}{\bibfnamefont{T.}~\bibnamefont{Oikawa}},
  \bibinfo{author}{\bibfnamefont{T.}~\bibnamefont{Shiraishi}},
  \bibinfo{author}{\bibfnamefont{T.}~\bibnamefont{Shimizu}},
  \bibinfo{author}{\bibfnamefont{H.}~\bibnamefont{Tanaka}},
  \bibinfo{author}{\bibfnamefont{N.}~\bibnamefont{Kanenko}},
  \bibinfo{author}{\bibfnamefont{R.}~\bibnamefont{Maran}},
  \bibinfo{author}{\bibfnamefont{T.}~\bibnamefont{Yamada}},
  \bibinfo{author}{\bibfnamefont{Y.}~\bibnamefont{Imai}}, \bibnamefont{et~al.},
  \bibinfo{journal}{Sci. Rep.} \textbf{\bibinfo{volume}{7}},
  \bibinfo{pages}{9641} (\bibinfo{year}{2017}).

\bibitem[{\citenamefont{Cheong and Mostovoy}(2007)}]{cheong_nmat_2007}
\bibinfo{author}{\bibfnamefont{S.-W.} \bibnamefont{Cheong}} \bibnamefont{and}
  \bibinfo{author}{\bibfnamefont{M.}~\bibnamefont{Mostovoy}},
  \bibinfo{journal}{Nat. Mater.} \textbf{\bibinfo{volume}{6}},
  \bibinfo{pages}{13} (\bibinfo{year}{2007}),
  \urlprefix\url{http://dx.doi.org/10.1038/nmat1804}.

\bibitem[{\citenamefont{Hoffmann et~al.}(2011)\citenamefont{Hoffmann, Thielen,
  Becker, Bohat\'y, and Fiebig}}]{hoffmann_prb_2011}
\bibinfo{author}{\bibfnamefont{T.}~\bibnamefont{Hoffmann}},
  \bibinfo{author}{\bibfnamefont{P.}~\bibnamefont{Thielen}},
  \bibinfo{author}{\bibfnamefont{P.}~\bibnamefont{Becker}},
  \bibinfo{author}{\bibfnamefont{L.}~\bibnamefont{Bohat\'y}}, \bibnamefont{and}
  \bibinfo{author}{\bibfnamefont{M.}~\bibnamefont{Fiebig}},
  \bibinfo{journal}{Phys. Rev. B} \textbf{\bibinfo{volume}{84}},
  \bibinfo{pages}{184404} (\bibinfo{year}{2011}),
  \urlprefix\url{https://link.aps.org/doi/10.1103/PhysRevB.84.184404}.

\bibitem[{\citenamefont{Kirilyuk et~al.}(2010)\citenamefont{Kirilyuk, Kimel,
  and Rasing}}]{kirilyuk_rmp_2010}
\bibinfo{author}{\bibfnamefont{A.}~\bibnamefont{Kirilyuk}},
  \bibinfo{author}{\bibfnamefont{A.~V.} \bibnamefont{Kimel}}, \bibnamefont{and}
  \bibinfo{author}{\bibfnamefont{T.}~\bibnamefont{Rasing}},
  \bibinfo{journal}{Rev. Mod. Phys.} \textbf{\bibinfo{volume}{82}},
  \bibinfo{pages}{2731} (\bibinfo{year}{2010}),
  \urlprefix\url{https://link.aps.org/doi/10.1103/RevModPhys.82.2731}.

\bibitem[{\citenamefont{Matsubara et~al.}(2015)\citenamefont{Matsubara,
  Schroer, Schmehl, Melville, Becher, Trujillo-Martinez, Schlom, Mannhart,
  Kroha, and Fiebig}}]{matsubara_natcomm_2015}
\bibinfo{author}{\bibfnamefont{M.}~\bibnamefont{Matsubara}},
  \bibinfo{author}{\bibfnamefont{A.}~\bibnamefont{Schroer}},
  \bibinfo{author}{\bibfnamefont{A.}~\bibnamefont{Schmehl}},
  \bibinfo{author}{\bibfnamefont{A.}~\bibnamefont{Melville}},
  \bibinfo{author}{\bibfnamefont{C.}~\bibnamefont{Becher}},
  \bibinfo{author}{\bibfnamefont{M.}~\bibnamefont{Trujillo-Martinez}},
  \bibinfo{author}{\bibfnamefont{D.~G.} \bibnamefont{Schlom}},
  \bibinfo{author}{\bibfnamefont{J.}~\bibnamefont{Mannhart}},
  \bibinfo{author}{\bibfnamefont{J.}~\bibnamefont{Kroha}}, \bibnamefont{and}
  \bibinfo{author}{\bibfnamefont{M.}~\bibnamefont{Fiebig}},
  \bibinfo{journal}{Nat. Comm.} \textbf{\bibinfo{volume}{6}},
  \bibinfo{pages}{6724} (\bibinfo{year}{2015}).

\bibitem[{\citenamefont{Johnson et~al.}(2012)\citenamefont{Johnson, de~Souza,
  Staub, Beaud, M\"ohr-Vorobeva, Ingold, Caviezel, Scagnoli, Schlotter, Turner
  et~al.}}]{johnson_prl_2012}
\bibinfo{author}{\bibfnamefont{S.~L.} \bibnamefont{Johnson}},
  \bibinfo{author}{\bibfnamefont{R.~A.} \bibnamefont{de~Souza}},
  \bibinfo{author}{\bibfnamefont{U.}~\bibnamefont{Staub}},
  \bibinfo{author}{\bibfnamefont{P.}~\bibnamefont{Beaud}},
  \bibinfo{author}{\bibfnamefont{E.}~\bibnamefont{M\"ohr-Vorobeva}},
  \bibinfo{author}{\bibfnamefont{G.}~\bibnamefont{Ingold}},
  \bibinfo{author}{\bibfnamefont{A.}~\bibnamefont{Caviezel}},
  \bibinfo{author}{\bibfnamefont{V.}~\bibnamefont{Scagnoli}},
  \bibinfo{author}{\bibfnamefont{W.~F.} \bibnamefont{Schlotter}},
  \bibinfo{author}{\bibfnamefont{J.~J.} \bibnamefont{Turner}},
  \bibnamefont{et~al.}, \bibinfo{journal}{Phys. Rev. Lett.}
  \textbf{\bibinfo{volume}{108}}, \bibinfo{pages}{037203}
  (\bibinfo{year}{2012}),
  \urlprefix\url{https://link.aps.org/doi/10.1103/PhysRevLett.108.037203}.

\bibitem[{\citenamefont{Kampfrath et~al.}(2011)\citenamefont{Kampfrath, Sell,
  Klatt, Pashkin, Maehrlein, Dekorsy, Wolf, Fiebig, Leitenstorfer, and
  Huber}}]{kampfrath_natph_2011}
\bibinfo{author}{\bibfnamefont{T.}~\bibnamefont{Kampfrath}},
  \bibinfo{author}{\bibfnamefont{A.}~\bibnamefont{Sell}},
  \bibinfo{author}{\bibfnamefont{G.}~\bibnamefont{Klatt}},
  \bibinfo{author}{\bibfnamefont{A.}~\bibnamefont{Pashkin}},
  \bibinfo{author}{\bibfnamefont{S.}~\bibnamefont{Maehrlein}},
  \bibinfo{author}{\bibfnamefont{T.}~\bibnamefont{Dekorsy}},
  \bibinfo{author}{\bibfnamefont{M.}~\bibnamefont{Wolf}},
  \bibinfo{author}{\bibfnamefont{M.}~\bibnamefont{Fiebig}},
  \bibinfo{author}{\bibfnamefont{A.}~\bibnamefont{Leitenstorfer}},
  \bibnamefont{and} \bibinfo{author}{\bibfnamefont{R.}~\bibnamefont{Huber}},
  \bibinfo{journal}{Nat. Phot.} \textbf{\bibinfo{volume}{5}},
  \bibinfo{pages}{31} (\bibinfo{year}{2011}).

\bibitem[{\citenamefont{Kubacka et~al.}(2014)\citenamefont{Kubacka, Johnson,
  Hoffmann, Vicario, de~Jong, Beaud, Gr{\"u}bel, Huang, Huber, Patthey
  et~al.}}]{kubacka_science_2014}
\bibinfo{author}{\bibfnamefont{T.}~\bibnamefont{Kubacka}},
  \bibinfo{author}{\bibfnamefont{J.~A.} \bibnamefont{Johnson}},
  \bibinfo{author}{\bibfnamefont{M.~C.} \bibnamefont{Hoffmann}},
  \bibinfo{author}{\bibfnamefont{C.}~\bibnamefont{Vicario}},
  \bibinfo{author}{\bibfnamefont{S.}~\bibnamefont{de~Jong}},
  \bibinfo{author}{\bibfnamefont{P.}~\bibnamefont{Beaud}},
  \bibinfo{author}{\bibfnamefont{S.}~\bibnamefont{Gr{\"u}bel}},
  \bibinfo{author}{\bibfnamefont{S.-W.} \bibnamefont{Huang}},
  \bibinfo{author}{\bibfnamefont{L.}~\bibnamefont{Huber}},
  \bibinfo{author}{\bibfnamefont{L.}~\bibnamefont{Patthey}},
  \bibnamefont{et~al.}, \bibinfo{journal}{Science}
  \textbf{\bibinfo{volume}{343}}, \bibinfo{pages}{1333} (\bibinfo{year}{2014}),
  \urlprefix\url{http://science.sciencemag.org/content/343/6177/1333}.

\bibitem[{\citenamefont{Kimel et~al.}(2007)\citenamefont{Kimel, Kirilyuk,
  Hansteen, Pisarev, and Rasing}}]{kimel_jpcm_2007}
\bibinfo{author}{\bibfnamefont{A.~V.} \bibnamefont{Kimel}},
  \bibinfo{author}{\bibfnamefont{A.}~\bibnamefont{Kirilyuk}},
  \bibinfo{author}{\bibfnamefont{F.}~\bibnamefont{Hansteen}},
  \bibinfo{author}{\bibfnamefont{R.~V.} \bibnamefont{Pisarev}},
  \bibnamefont{and} \bibinfo{author}{\bibfnamefont{T.}~\bibnamefont{Rasing}},
  \bibinfo{journal}{J. Phys.: Condens. Matter} \textbf{\bibinfo{volume}{19}},
  \bibinfo{pages}{043201} (\bibinfo{year}{2007}),
  \urlprefix\url{http://stacks.iop.org/0953-8984/19/i=4/a=043201}.

\bibitem[{\citenamefont{Lambert et~al.}(2014)\citenamefont{Lambert, Mangin,
  Varaprasad, Takahashi, Hehn, Cinchetti, Malinowski, Hono, Fainman,
  Aeschlimann et~al.}}]{lambert_science_2014}
\bibinfo{author}{\bibfnamefont{C.-H.} \bibnamefont{Lambert}},
  \bibinfo{author}{\bibfnamefont{S.}~\bibnamefont{Mangin}},
  \bibinfo{author}{\bibfnamefont{B.~S. D. C.~S.} \bibnamefont{Varaprasad}},
  \bibinfo{author}{\bibfnamefont{Y.~K.} \bibnamefont{Takahashi}},
  \bibinfo{author}{\bibfnamefont{M.}~\bibnamefont{Hehn}},
  \bibinfo{author}{\bibfnamefont{M.}~\bibnamefont{Cinchetti}},
  \bibinfo{author}{\bibfnamefont{G.}~\bibnamefont{Malinowski}},
  \bibinfo{author}{\bibfnamefont{K.}~\bibnamefont{Hono}},
  \bibinfo{author}{\bibfnamefont{Y.}~\bibnamefont{Fainman}},
  \bibinfo{author}{\bibfnamefont{M.}~\bibnamefont{Aeschlimann}},
  \bibnamefont{et~al.}, \bibinfo{journal}{Science}
  \textbf{\bibinfo{volume}{345}}, \bibinfo{pages}{1337} (\bibinfo{year}{2014}),
  \urlprefix\url{http://science.sciencemag.org/content/345/6202/1337}.

\bibitem[{\citenamefont{Tokura and Kida}(2011)}]{tokura_phil_2011}
\bibinfo{author}{\bibfnamefont{Y.}~\bibnamefont{Tokura}} \bibnamefont{and}
  \bibinfo{author}{\bibfnamefont{N.}~\bibnamefont{Kida}},
  \bibinfo{journal}{Phil. Trans. Royal Soc. A} \textbf{\bibinfo{volume}{369}},
  \bibinfo{pages}{3679} (\bibinfo{year}{2011}),
  \urlprefix\url{http://rsta.royalsocietypublishing.org/content/369/1951/3679.abstract}.

\bibitem[{\citenamefont{Pimenov et~al.}(2006)\citenamefont{Pimenov, Mukhin,
  Ivanov, Travkin, Balbashov, and Loidl}}]{pimenov_nphys_2006}
\bibinfo{author}{\bibfnamefont{A.}~\bibnamefont{Pimenov}},
  \bibinfo{author}{\bibfnamefont{A.~A.} \bibnamefont{Mukhin}},
  \bibinfo{author}{\bibfnamefont{V.~Y.} \bibnamefont{Ivanov}},
  \bibinfo{author}{\bibfnamefont{V.~D.} \bibnamefont{Travkin}},
  \bibinfo{author}{\bibfnamefont{A.~M.} \bibnamefont{Balbashov}},
  \bibnamefont{and} \bibinfo{author}{\bibfnamefont{A.}~\bibnamefont{Loidl}},
  \bibinfo{journal}{Nat. Phys.} \textbf{\bibinfo{volume}{2}},
  \bibinfo{pages}{97} (\bibinfo{year}{2006}),
  \urlprefix\url{http://dx.doi.org/10.1038/nphys212}.

\bibitem[{\citenamefont{Sushkov et~al.}(2007)\citenamefont{Sushkov, Aguilar,
  Park, Cheong, and Drew}}]{sushkov_prl_2007}
\bibinfo{author}{\bibfnamefont{A.~B.} \bibnamefont{Sushkov}},
  \bibinfo{author}{\bibfnamefont{R.~V.} \bibnamefont{Aguilar}},
  \bibinfo{author}{\bibfnamefont{S.}~\bibnamefont{Park}},
  \bibinfo{author}{\bibfnamefont{S.-W.} \bibnamefont{Cheong}},
  \bibnamefont{and} \bibinfo{author}{\bibfnamefont{H.~D.} \bibnamefont{Drew}},
  \bibinfo{journal}{Phys. Rev. Lett.} \textbf{\bibinfo{volume}{98}},
  \bibinfo{eid}{027202} (\bibinfo{year}{2007}),
  \urlprefix\url{http://link.aps.org/abstract/PRL/v98/e027202}.

\bibitem[{\citenamefont{Kida et~al.}(2009)\citenamefont{Kida, Takahashi, Lee,
  Shimano, Yamasaki, Kaneko, Miyahara, Furukawa, Arima, and
  Tokura}}]{kida_josab_2009}
\bibinfo{author}{\bibfnamefont{N.}~\bibnamefont{Kida}},
  \bibinfo{author}{\bibfnamefont{Y.}~\bibnamefont{Takahashi}},
  \bibinfo{author}{\bibfnamefont{J.~S.} \bibnamefont{Lee}},
  \bibinfo{author}{\bibfnamefont{R.}~\bibnamefont{Shimano}},
  \bibinfo{author}{\bibfnamefont{Y.}~\bibnamefont{Yamasaki}},
  \bibinfo{author}{\bibfnamefont{Y.}~\bibnamefont{Kaneko}},
  \bibinfo{author}{\bibfnamefont{S.}~\bibnamefont{Miyahara}},
  \bibinfo{author}{\bibfnamefont{N.}~\bibnamefont{Furukawa}},
  \bibinfo{author}{\bibfnamefont{T.}~\bibnamefont{Arima}}, \bibnamefont{and}
  \bibinfo{author}{\bibfnamefont{Y.}~\bibnamefont{Tokura}},
  \bibinfo{journal}{J. Opt. Soc. Am. B} \textbf{\bibinfo{volume}{26}},
  \bibinfo{pages}{A35} (\bibinfo{year}{2009}),
  \urlprefix\url{http://josab.osa.org/abstract.cfm?URI=josab-26-9-A35}.

\bibitem[{\citenamefont{Rovillain et~al.}(2010)\citenamefont{Rovillain,
  de~Sousa, Gallais, Sacuto, Measson, Colson, Forget, Bibes, Barthelemy, and
  Cazayous}}]{rovillain_nmat_2010}
\bibinfo{author}{\bibfnamefont{P.}~\bibnamefont{Rovillain}},
  \bibinfo{author}{\bibfnamefont{R.}~\bibnamefont{de~Sousa}},
  \bibinfo{author}{\bibfnamefont{Y.}~\bibnamefont{Gallais}},
  \bibinfo{author}{\bibfnamefont{A.}~\bibnamefont{Sacuto}},
  \bibinfo{author}{\bibfnamefont{M.~A.} \bibnamefont{Measson}},
  \bibinfo{author}{\bibfnamefont{D.}~\bibnamefont{Colson}},
  \bibinfo{author}{\bibfnamefont{A.}~\bibnamefont{Forget}},
  \bibinfo{author}{\bibfnamefont{M.}~\bibnamefont{Bibes}},
  \bibinfo{author}{\bibfnamefont{A.}~\bibnamefont{Barthelemy}},
  \bibnamefont{and} \bibinfo{author}{\bibfnamefont{M.}~\bibnamefont{Cazayous}},
  \bibinfo{journal}{Nat. Mater.} \textbf{\bibinfo{volume}{9}},
  \bibinfo{pages}{975} (\bibinfo{year}{2010}).

\bibitem[{\citenamefont{Shuvaev et~al.}(2013)\citenamefont{Shuvaev, Dziom,
  Pimenov, Schiebl, Mukhin, Komarek, Finger, Braden, and
  Pimenov}}]{shuvaev_prl_2013}
\bibinfo{author}{\bibfnamefont{A.}~\bibnamefont{Shuvaev}},
  \bibinfo{author}{\bibfnamefont{V.}~\bibnamefont{Dziom}},
  \bibinfo{author}{\bibfnamefont{A.}~\bibnamefont{Pimenov}},
  \bibinfo{author}{\bibfnamefont{M.}~\bibnamefont{Schiebl}},
  \bibinfo{author}{\bibfnamefont{A.~A.} \bibnamefont{Mukhin}},
  \bibinfo{author}{\bibfnamefont{A.~C.} \bibnamefont{Komarek}},
  \bibinfo{author}{\bibfnamefont{T.}~\bibnamefont{Finger}},
  \bibinfo{author}{\bibfnamefont{M.}~\bibnamefont{Braden}}, \bibnamefont{and}
  \bibinfo{author}{\bibfnamefont{A.}~\bibnamefont{Pimenov}},
  \bibinfo{journal}{Phys. Rev. Lett.} \textbf{\bibinfo{volume}{111}},
  \bibinfo{pages}{227201} (\bibinfo{year}{2013}),
  \urlprefix\url{http://link.aps.org/doi/10.1103/PhysRevLett.111.227201}.

\bibitem[{\citenamefont{Shastry et~al.}(2004)\citenamefont{Shastry, Srinivasan,
  Bichurin, Petrov, and Tatarenko}}]{shastry_prb_2004}
\bibinfo{author}{\bibfnamefont{S.}~\bibnamefont{Shastry}},
  \bibinfo{author}{\bibfnamefont{G.}~\bibnamefont{Srinivasan}},
  \bibinfo{author}{\bibfnamefont{M.~I.} \bibnamefont{Bichurin}},
  \bibinfo{author}{\bibfnamefont{V.~M.} \bibnamefont{Petrov}},
  \bibnamefont{and} \bibinfo{author}{\bibfnamefont{A.~S.}
  \bibnamefont{Tatarenko}}, \bibinfo{journal}{Phys. Rev. B}
  \textbf{\bibinfo{volume}{70}}, \bibinfo{pages}{064416}
  (\bibinfo{year}{2004}),
  \urlprefix\url{https://link.aps.org/doi/10.1103/PhysRevB.70.064416}.

\bibitem[{\citenamefont{Das et~al.}(2009)\citenamefont{Das, Song, Mo, Krivosik,
  and Patton}}]{das_advmat_2009}
\bibinfo{author}{\bibfnamefont{J.}~\bibnamefont{Das}},
  \bibinfo{author}{\bibfnamefont{Y.-Y.} \bibnamefont{Song}},
  \bibinfo{author}{\bibfnamefont{N.}~\bibnamefont{Mo}},
  \bibinfo{author}{\bibfnamefont{P.}~\bibnamefont{Krivosik}}, \bibnamefont{and}
  \bibinfo{author}{\bibfnamefont{C.~E.} \bibnamefont{Patton}},
  \bibinfo{journal}{Adv. Mater.} \textbf{\bibinfo{volume}{21}},
  \bibinfo{pages}{2045} (\bibinfo{year}{2009}),
  \urlprefix\url{http://dx.doi.org/10.1002/adma.200803376}.

\bibitem[{\citenamefont{Zhu et~al.}(2012)\citenamefont{Zhu, Katine, Rowlands,
  Chen, Duan, Alzate, Upadhyaya, Langer, Amiri, Wang et~al.}}]{zhu_prl_2012}
\bibinfo{author}{\bibfnamefont{J.}~\bibnamefont{Zhu}},
  \bibinfo{author}{\bibfnamefont{J.~A.} \bibnamefont{Katine}},
  \bibinfo{author}{\bibfnamefont{G.~E.} \bibnamefont{Rowlands}},
  \bibinfo{author}{\bibfnamefont{Y.-J.} \bibnamefont{Chen}},
  \bibinfo{author}{\bibfnamefont{Z.}~\bibnamefont{Duan}},
  \bibinfo{author}{\bibfnamefont{J.~G.} \bibnamefont{Alzate}},
  \bibinfo{author}{\bibfnamefont{P.}~\bibnamefont{Upadhyaya}},
  \bibinfo{author}{\bibfnamefont{J.}~\bibnamefont{Langer}},
  \bibinfo{author}{\bibfnamefont{P.~K.} \bibnamefont{Amiri}},
  \bibinfo{author}{\bibfnamefont{K.~L.} \bibnamefont{Wang}},
  \bibnamefont{et~al.}, \bibinfo{journal}{Phys. Rev. Lett.}
  \textbf{\bibinfo{volume}{108}}, \bibinfo{pages}{197203}
  (\bibinfo{year}{2012}),
  \urlprefix\url{https://link.aps.org/doi/10.1103/PhysRevLett.108.197203}.

\bibitem[{\citenamefont{Nozaki et~al.}(2012)\citenamefont{Nozaki, Shiota, Miwa,
  Murakami, Bonell, Ishibashi, Kubota, Yakushiji, Saruya, Fukushima
  et~al.}}]{nozaki_natphys_2012}
\bibinfo{author}{\bibfnamefont{T.}~\bibnamefont{Nozaki}},
  \bibinfo{author}{\bibfnamefont{Y.}~\bibnamefont{Shiota}},
  \bibinfo{author}{\bibfnamefont{S.}~\bibnamefont{Miwa}},
  \bibinfo{author}{\bibfnamefont{S.}~\bibnamefont{Murakami}},
  \bibinfo{author}{\bibfnamefont{F.}~\bibnamefont{Bonell}},
  \bibinfo{author}{\bibfnamefont{S.}~\bibnamefont{Ishibashi}},
  \bibinfo{author}{\bibfnamefont{H.}~\bibnamefont{Kubota}},
  \bibinfo{author}{\bibfnamefont{K.}~\bibnamefont{Yakushiji}},
  \bibinfo{author}{\bibfnamefont{T.}~\bibnamefont{Saruya}},
  \bibinfo{author}{\bibfnamefont{A.}~\bibnamefont{Fukushima}},
  \bibnamefont{et~al.}, \bibinfo{journal}{Nat. Phys.}
  \textbf{\bibinfo{volume}{8}}, \bibinfo{pages}{491} (\bibinfo{year}{2012}).

\bibitem[{\citenamefont{Vasiliev and Popova}(2006)}]{vasiliev_ltp_2006}
\bibinfo{author}{\bibfnamefont{A.~N.} \bibnamefont{Vasiliev}} \bibnamefont{and}
  \bibinfo{author}{\bibfnamefont{E.~A.} \bibnamefont{Popova}},
  \bibinfo{journal}{Low Temp. Phys.} \textbf{\bibinfo{volume}{32}},
  \bibinfo{pages}{735} (\bibinfo{year}{2006}),
  \urlprefix\url{http://link.aip.org/link/?LTP/32/735/1}.

\bibitem[{\citenamefont{Kadomtseva et~al.}(2010)\citenamefont{Kadomtseva,
  Popov, Vorob'ev, Pyatakov, Krotov, Kamilov, Ivanov, Mukhin, Zvezdin,
  Kuz'menko et~al.}}]{kadomtseva_ltp_2010}
\bibinfo{author}{\bibfnamefont{A.~M.} \bibnamefont{Kadomtseva}},
  \bibinfo{author}{\bibfnamefont{Y.~F.} \bibnamefont{Popov}},
  \bibinfo{author}{\bibfnamefont{G.~P.} \bibnamefont{Vorob'ev}},
  \bibinfo{author}{\bibfnamefont{A.~P.} \bibnamefont{Pyatakov}},
  \bibinfo{author}{\bibfnamefont{S.~S.} \bibnamefont{Krotov}},
  \bibinfo{author}{\bibfnamefont{K.~I.} \bibnamefont{Kamilov}},
  \bibinfo{author}{\bibfnamefont{V.~Y.} \bibnamefont{Ivanov}},
  \bibinfo{author}{\bibfnamefont{A.~A.} \bibnamefont{Mukhin}},
  \bibinfo{author}{\bibfnamefont{A.~K.} \bibnamefont{Zvezdin}},
  \bibinfo{author}{\bibfnamefont{A.~M.} \bibnamefont{Kuz'menko}},
  \bibnamefont{et~al.}, \bibinfo{journal}{Low Temp. Phys.}
  \textbf{\bibinfo{volume}{36}}, \bibinfo{pages}{511} (\bibinfo{year}{2010}),
  \urlprefix\url{http://link.aip.org/link/?LTP/36/511/1}.

\bibitem[{\citenamefont{Kadomtseva et~al.}(2012)\citenamefont{Kadomtseva,
  Vorob'ev, Popov, Pyatakov, Mukhin, Ivanov, Zvezdin, Gudim, Temerov, and
  Bezmaternykh}}]{kadomtseva_jetp_2012}
\bibinfo{author}{\bibfnamefont{A.~M.} \bibnamefont{Kadomtseva}},
  \bibinfo{author}{\bibfnamefont{G.~P.} \bibnamefont{Vorob'ev}},
  \bibinfo{author}{\bibfnamefont{Y.~F.} \bibnamefont{Popov}},
  \bibinfo{author}{\bibfnamefont{A.~P.} \bibnamefont{Pyatakov}},
  \bibinfo{author}{\bibfnamefont{A.~A.} \bibnamefont{Mukhin}},
  \bibinfo{author}{\bibfnamefont{V.~Y.} \bibnamefont{Ivanov}},
  \bibinfo{author}{\bibfnamefont{A.~K.} \bibnamefont{Zvezdin}},
  \bibinfo{author}{\bibfnamefont{I.~A.} \bibnamefont{Gudim}},
  \bibinfo{author}{\bibfnamefont{V.~L.} \bibnamefont{Temerov}},
  \bibnamefont{and} \bibinfo{author}{\bibfnamefont{L.~N.}
  \bibnamefont{Bezmaternykh}}, \bibinfo{journal}{J. Exp. Theor. Phys.}
  \textbf{\bibinfo{volume}{114}}, \bibinfo{pages}{810} (\bibinfo{year}{2012}),
  \urlprefix\url{http://dx.doi.org/10.1134/S1063776112030053}.

\bibitem[{\citenamefont{Camp\'{a} et~al.}(1997)\citenamefont{Camp\'{a},
  Cascales, Gutiérrez-Puebla, Monge, Rasines, and
  Ruíz-Valero}}]{campa_chemm_1997}
\bibinfo{author}{\bibfnamefont{J.~A.} \bibnamefont{Camp\'{a}}},
  \bibinfo{author}{\bibfnamefont{C.}~\bibnamefont{Cascales}},
  \bibinfo{author}{\bibfnamefont{E.}~\bibnamefont{Gutiérrez-Puebla}},
  \bibinfo{author}{\bibfnamefont{M.~A.} \bibnamefont{Monge}},
  \bibinfo{author}{\bibfnamefont{I.}~\bibnamefont{Rasines}}, \bibnamefont{and}
  \bibinfo{author}{\bibfnamefont{C.}~\bibnamefont{Ruíz-Valero}},
  \bibinfo{journal}{Chem. Mater.} \textbf{\bibinfo{volume}{9}},
  \bibinfo{pages}{237} (\bibinfo{year}{1997}),
  \urlprefix\url{http://dx.doi.org/10.1021/cm960313m}.

\bibitem[{\citenamefont{Hinatsu et~al.}(2003)\citenamefont{Hinatsu, Doi, Ito,
  Wakeshima, and Alemi}}]{hinatsu_jssch_2002}
\bibinfo{author}{\bibfnamefont{Y.}~\bibnamefont{Hinatsu}},
  \bibinfo{author}{\bibfnamefont{Y.}~\bibnamefont{Doi}},
  \bibinfo{author}{\bibfnamefont{K.}~\bibnamefont{Ito}},
  \bibinfo{author}{\bibfnamefont{M.}~\bibnamefont{Wakeshima}},
  \bibnamefont{and} \bibinfo{author}{\bibfnamefont{A.}~\bibnamefont{Alemi}},
  \bibinfo{journal}{J. Solid State Chem.} \textbf{\bibinfo{volume}{172}},
  \bibinfo{pages}{438 } (\bibinfo{year}{2003}),
  \urlprefix\url{http://www.sciencedirect.com/science/article/pii/S0022459603000288}.

\bibitem[{\citenamefont{Fausti et~al.}(2006)\citenamefont{Fausti, Nugroho, van
  Loosdrecht, Klimin, Popova, and Bezmaternykh}}]{fausti_prb_2006}
\bibinfo{author}{\bibfnamefont{D.}~\bibnamefont{Fausti}},
  \bibinfo{author}{\bibfnamefont{A.~A.} \bibnamefont{Nugroho}},
  \bibinfo{author}{\bibfnamefont{P.~H.~M.} \bibnamefont{van Loosdrecht}},
  \bibinfo{author}{\bibfnamefont{S.~A.} \bibnamefont{Klimin}},
  \bibinfo{author}{\bibfnamefont{M.~N.} \bibnamefont{Popova}},
  \bibnamefont{and} \bibinfo{author}{\bibfnamefont{L.~N.}
  \bibnamefont{Bezmaternykh}}, \bibinfo{journal}{Phys. Rev. B}
  \textbf{\bibinfo{volume}{74}}, \bibinfo{pages}{024403}
  (\bibinfo{year}{2006}),
  \urlprefix\url{https://link.aps.org/doi/10.1103/PhysRevB.74.024403}.

\bibitem[{\citenamefont{Popova}(2009)}]{popova_jre_2009}
\bibinfo{author}{\bibfnamefont{M.~N.} \bibnamefont{Popova}},
  \bibinfo{journal}{J. Rare Earths} \textbf{\bibinfo{volume}{27}},
  \bibinfo{pages}{607 } (\bibinfo{year}{2009}),
  \urlprefix\url{http://www.sciencedirect.com/science/article/pii/S1002072108602987}.

\bibitem[{sup()}]{suppl}
\bibinfo{note}{See the Supplemental Material for detail}.

\bibitem[{\citenamefont{Zvezdin et~al.}(2006)\citenamefont{Zvezdin, Vorob'ev,
  Kadomtseva, Popov, Pyatakov, Bezmaternykh, Kuvardin, and
  Popova}}]{zvezdin_jetpl_2006}
\bibinfo{author}{\bibfnamefont{A.~K.} \bibnamefont{Zvezdin}},
  \bibinfo{author}{\bibfnamefont{G.~P.} \bibnamefont{Vorob'ev}},
  \bibinfo{author}{\bibfnamefont{A.~M.} \bibnamefont{Kadomtseva}},
  \bibinfo{author}{\bibfnamefont{Y.~F.} \bibnamefont{Popov}},
  \bibinfo{author}{\bibfnamefont{A.~P.} \bibnamefont{Pyatakov}},
  \bibinfo{author}{\bibfnamefont{L.~N.} \bibnamefont{Bezmaternykh}},
  \bibinfo{author}{\bibfnamefont{A.~V.} \bibnamefont{Kuvardin}},
  \bibnamefont{and} \bibinfo{author}{\bibfnamefont{E.~A.}
  \bibnamefont{Popova}}, \bibinfo{journal}{JETP Lett.}
  \textbf{\bibinfo{volume}{83}}, \bibinfo{pages}{509} (\bibinfo{year}{2006}),
  \urlprefix\url{http://dx.doi.org/10.1134/S0021364006110099}.

\bibitem[{\citenamefont{Popov et~al.}(2010)\citenamefont{Popov, Pyatakov,
  Kadomtseva, Vorob'ev, Zvezdin, Mukhin, Ivanov, and Gudim}}]{popov_jetp_2010}
\bibinfo{author}{\bibfnamefont{Y.~F.} \bibnamefont{Popov}},
  \bibinfo{author}{\bibfnamefont{A.~P.} \bibnamefont{Pyatakov}},
  \bibinfo{author}{\bibfnamefont{A.~M.} \bibnamefont{Kadomtseva}},
  \bibinfo{author}{\bibfnamefont{G.~P.} \bibnamefont{Vorob'ev}},
  \bibinfo{author}{\bibfnamefont{A.~K.} \bibnamefont{Zvezdin}},
  \bibinfo{author}{\bibfnamefont{A.~A.} \bibnamefont{Mukhin}},
  \bibinfo{author}{\bibfnamefont{V.~Y.} \bibnamefont{Ivanov}},
  \bibnamefont{and} \bibinfo{author}{\bibfnamefont{I.~A.} \bibnamefont{Gudim}},
  \bibinfo{journal}{J. Exp. Theor. Phys.} \textbf{\bibinfo{volume}{111}},
  \bibinfo{pages}{199} (\bibinfo{year}{2010}),
  \urlprefix\url{http://dx.doi.org/10.1134/S1063776110080066}.

\bibitem[{\citenamefont{Mukhin et~al.}(2011)\citenamefont{Mukhin, Vorob'ev,
  Ivanov, Kadomtseva, Narizhnaya, Kuz'menko, Popov, Bezmaternykh, and
  Gudim}}]{mukhin_jetpl_2011}
\bibinfo{author}{\bibfnamefont{A.~A.} \bibnamefont{Mukhin}},
  \bibinfo{author}{\bibfnamefont{G.~P.} \bibnamefont{Vorob'ev}},
  \bibinfo{author}{\bibfnamefont{V.~Y.} \bibnamefont{Ivanov}},
  \bibinfo{author}{\bibfnamefont{A.~M.} \bibnamefont{Kadomtseva}},
  \bibinfo{author}{\bibfnamefont{A.~S.} \bibnamefont{Narizhnaya}},
  \bibinfo{author}{\bibfnamefont{A.~M.} \bibnamefont{Kuz'menko}},
  \bibinfo{author}{\bibfnamefont{Y.~F.} \bibnamefont{Popov}},
  \bibinfo{author}{\bibfnamefont{L.~N.} \bibnamefont{Bezmaternykh}},
  \bibnamefont{and} \bibinfo{author}{\bibfnamefont{I.~A.} \bibnamefont{Gudim}},
  \bibinfo{journal}{JETP Lett.} \textbf{\bibinfo{volume}{93}},
  \bibinfo{pages}{275} (\bibinfo{year}{2011}),
  \urlprefix\url{http://dx.doi.org/10.1134/S0021364011050079}.

\bibitem[{\citenamefont{Ritter et~al.}(2012)\citenamefont{Ritter, Pankrats,
  Gudim, and Vorotynov}}]{ritter_jpcm_2012}
\bibinfo{author}{\bibfnamefont{C.}~\bibnamefont{Ritter}},
  \bibinfo{author}{\bibfnamefont{A.}~\bibnamefont{Pankrats}},
  \bibinfo{author}{\bibfnamefont{I.}~\bibnamefont{Gudim}}, \bibnamefont{and}
  \bibinfo{author}{\bibfnamefont{A.}~\bibnamefont{Vorotynov}},
  \bibinfo{journal}{J. Phys.: Condens. Matter} \textbf{\bibinfo{volume}{24}},
  \bibinfo{pages}{386002} (\bibinfo{year}{2012}),
  \urlprefix\url{http://stacks.iop.org/0953-8984/24/i=38/a=386002}.

\bibitem[{\citenamefont{Popov et~al.}(2013)\citenamefont{Popov, Plokhov, and
  Zvezdin}}]{popov_prb_2013}
\bibinfo{author}{\bibfnamefont{A.~I.} \bibnamefont{Popov}},
  \bibinfo{author}{\bibfnamefont{D.~I.} \bibnamefont{Plokhov}},
  \bibnamefont{and} \bibinfo{author}{\bibfnamefont{A.~K.}
  \bibnamefont{Zvezdin}}, \bibinfo{journal}{Phys. Rev. B}
  \textbf{\bibinfo{volume}{87}}, \bibinfo{pages}{024413}
  (\bibinfo{year}{2013}),
  \urlprefix\url{http://link.aps.org/doi/10.1103/PhysRevB.87.024413}.

\bibitem[{\citenamefont{O'Dell}(1970)}]{dell_book}
\bibinfo{author}{\bibfnamefont{T.~H.} \bibnamefont{O'Dell}},
  \emph{\bibinfo{title}{The electrodynamics of magneto-electric media}}
  (\bibinfo{publisher}{North-Holland Publishing}, \bibinfo{address}{Amsterdam},
  \bibinfo{year}{1970}).

\bibitem[{\citenamefont{Freidman et~al.}(2015)\citenamefont{Freidman, Balaev,
  Dubrovskii, Eremin, Shaikhutdinov, Temerov, and Gudim}}]{freidman_ftt_2015}
\bibinfo{author}{\bibfnamefont{A.~L.} \bibnamefont{Freidman}},
  \bibinfo{author}{\bibfnamefont{A.~D.} \bibnamefont{Balaev}},
  \bibinfo{author}{\bibfnamefont{A.~A.} \bibnamefont{Dubrovskii}},
  \bibinfo{author}{\bibfnamefont{E.~V.} \bibnamefont{Eremin}},
  \bibinfo{author}{\bibfnamefont{K.~A.} \bibnamefont{Shaikhutdinov}},
  \bibinfo{author}{\bibfnamefont{V.~L.} \bibnamefont{Temerov}},
  \bibnamefont{and} \bibinfo{author}{\bibfnamefont{I.~A.} \bibnamefont{Gudim}},
  \bibinfo{journal}{Phys. Sol. State} \textbf{\bibinfo{volume}{57}},
  \bibinfo{pages}{1357} (\bibinfo{year}{2015}),
  \urlprefix\url{http://dx.doi.org/10.1134/S1063783415070112}.

\bibitem[{\citenamefont{Partzsch et~al.}(2016)\citenamefont{Partzsch,
  Hamann-Borrero, Mazzoli, Herrero-Martin, Valencia, Feyerherm, Dudzik,
  Vasiliev, Bezmaternykh, B\"uchner et~al.}}]{partzsch_prb_2016}
\bibinfo{author}{\bibfnamefont{S.}~\bibnamefont{Partzsch}},
  \bibinfo{author}{\bibfnamefont{J.-E.} \bibnamefont{Hamann-Borrero}},
  \bibinfo{author}{\bibfnamefont{C.}~\bibnamefont{Mazzoli}},
  \bibinfo{author}{\bibfnamefont{J.}~\bibnamefont{Herrero-Martin}},
  \bibinfo{author}{\bibfnamefont{S.}~\bibnamefont{Valencia}},
  \bibinfo{author}{\bibfnamefont{R.}~\bibnamefont{Feyerherm}},
  \bibinfo{author}{\bibfnamefont{E.}~\bibnamefont{Dudzik}},
  \bibinfo{author}{\bibfnamefont{A.}~\bibnamefont{Vasiliev}},
  \bibinfo{author}{\bibfnamefont{L.}~\bibnamefont{Bezmaternykh}},
  \bibinfo{author}{\bibfnamefont{B.}~\bibnamefont{B\"uchner}},
  \bibnamefont{et~al.}, \bibinfo{journal}{Phys. Rev. B}
  \textbf{\bibinfo{volume}{94}}, \bibinfo{pages}{054421}
  (\bibinfo{year}{2016}),
  \urlprefix\url{http://link.aps.org/doi/10.1103/PhysRevB.94.054421}.

\bibitem[{\citenamefont{Kuz'menko
  et~al.}(2011{\natexlab{a}})\citenamefont{Kuz'menko, Mukhin, Ivanov,
  Kadomtseva, and Bezmaternykh}}]{kuzmenko_jetpl_2011}
\bibinfo{author}{\bibfnamefont{A.~M.} \bibnamefont{Kuz'menko}},
  \bibinfo{author}{\bibfnamefont{A.~A.} \bibnamefont{Mukhin}},
  \bibinfo{author}{\bibfnamefont{V.~Y.} \bibnamefont{Ivanov}},
  \bibinfo{author}{\bibfnamefont{A.~M.} \bibnamefont{Kadomtseva}},
  \bibnamefont{and} \bibinfo{author}{\bibfnamefont{L.~N.}
  \bibnamefont{Bezmaternykh}}, \bibinfo{journal}{JETP Lett.}
  \textbf{\bibinfo{volume}{94}}, \bibinfo{pages}{294}
  (\bibinfo{year}{2011}{\natexlab{a}}),
  \urlprefix\url{http://dx.doi.org/10.1134/S0021364011160119}.

\bibitem[{\citenamefont{Kuz'menko
  et~al.}(2011{\natexlab{b}})\citenamefont{Kuz'menko, Mukhin, Ivanov,
  Kadomtseva, Lebedev, and Bezmaternykh}}]{kuzmenko_jetp_2011}
\bibinfo{author}{\bibfnamefont{A.~M.} \bibnamefont{Kuz'menko}},
  \bibinfo{author}{\bibfnamefont{A.~A.} \bibnamefont{Mukhin}},
  \bibinfo{author}{\bibfnamefont{V.~Y.} \bibnamefont{Ivanov}},
  \bibinfo{author}{\bibfnamefont{A.~M.} \bibnamefont{Kadomtseva}},
  \bibinfo{author}{\bibfnamefont{S.~P.} \bibnamefont{Lebedev}},
  \bibnamefont{and} \bibinfo{author}{\bibfnamefont{L.~N.}
  \bibnamefont{Bezmaternykh}}, \bibinfo{journal}{J. Exp. Theor. Phys.}
  \textbf{\bibinfo{volume}{113}}, \bibinfo{pages}{113}
  (\bibinfo{year}{2011}{\natexlab{b}}),
  \urlprefix\url{http://dx.doi.org/10.1134/S106377611105013X}.

\bibitem[{\citenamefont{Kuzmenko et~al.}(2014)\citenamefont{Kuzmenko, Shuvaev,
  Dziom, Pimenov, Schiebl, Mukhin, Ivanov, Bezmaternykh, and
  Pimenov}}]{kuzmenko_prb_2014}
\bibinfo{author}{\bibfnamefont{A.~M.} \bibnamefont{Kuzmenko}},
  \bibinfo{author}{\bibfnamefont{A.}~\bibnamefont{Shuvaev}},
  \bibinfo{author}{\bibfnamefont{V.}~\bibnamefont{Dziom}},
  \bibinfo{author}{\bibfnamefont{A.}~\bibnamefont{Pimenov}},
  \bibinfo{author}{\bibfnamefont{M.}~\bibnamefont{Schiebl}},
  \bibinfo{author}{\bibfnamefont{A.~A.} \bibnamefont{Mukhin}},
  \bibinfo{author}{\bibfnamefont{V.~Y.} \bibnamefont{Ivanov}},
  \bibinfo{author}{\bibfnamefont{L.~N.} \bibnamefont{Bezmaternykh}},
  \bibnamefont{and} \bibinfo{author}{\bibfnamefont{A.}~\bibnamefont{Pimenov}},
  \bibinfo{journal}{Phys. Rev. B} \textbf{\bibinfo{volume}{89}},
  \bibinfo{pages}{174407} (\bibinfo{year}{2014}),
  \urlprefix\url{http://link.aps.org/doi/10.1103/PhysRevB.89.174407}.

\bibitem[{\citenamefont{Kuzmenko et~al.}(2015)\citenamefont{Kuzmenko, Dziom,
  Shuvaev, Pimenov, Schiebl, Mukhin, Ivanov, Gudim, Bezmaternykh, and
  Pimenov}}]{kuzmenko_prb_2015}
\bibinfo{author}{\bibfnamefont{A.~M.} \bibnamefont{Kuzmenko}},
  \bibinfo{author}{\bibfnamefont{V.}~\bibnamefont{Dziom}},
  \bibinfo{author}{\bibfnamefont{A.}~\bibnamefont{Shuvaev}},
  \bibinfo{author}{\bibfnamefont{A.}~\bibnamefont{Pimenov}},
  \bibinfo{author}{\bibfnamefont{M.}~\bibnamefont{Schiebl}},
  \bibinfo{author}{\bibfnamefont{A.~A.} \bibnamefont{Mukhin}},
  \bibinfo{author}{\bibfnamefont{V.~Y.} \bibnamefont{Ivanov}},
  \bibinfo{author}{\bibfnamefont{I.~A.} \bibnamefont{Gudim}},
  \bibinfo{author}{\bibfnamefont{L.~N.} \bibnamefont{Bezmaternykh}},
  \bibnamefont{and} \bibinfo{author}{\bibfnamefont{A.}~\bibnamefont{Pimenov}},
  \bibinfo{journal}{Phys. Rev. B} \textbf{\bibinfo{volume}{92}},
  \bibinfo{pages}{184409} (\bibinfo{year}{2015}),
  \urlprefix\url{http://link.aps.org/doi/10.1103/PhysRevB.92.184409}.

\bibitem[{\citenamefont{Popova et~al.}(2007)\citenamefont{Popova, Chukalina,
  Stanislavchuk, Malkin, Zakirov, Antic-Fidancev, Popova, Bezmaternykh, and
  Temerov}}]{popova_prb_2007}
\bibinfo{author}{\bibfnamefont{M.~N.} \bibnamefont{Popova}},
  \bibinfo{author}{\bibfnamefont{E.~P.} \bibnamefont{Chukalina}},
  \bibinfo{author}{\bibfnamefont{T.~N.} \bibnamefont{Stanislavchuk}},
  \bibinfo{author}{\bibfnamefont{B.~Z.} \bibnamefont{Malkin}},
  \bibinfo{author}{\bibfnamefont{A.~R.} \bibnamefont{Zakirov}},
  \bibinfo{author}{\bibfnamefont{E.}~\bibnamefont{Antic-Fidancev}},
  \bibinfo{author}{\bibfnamefont{E.~A.} \bibnamefont{Popova}},
  \bibinfo{author}{\bibfnamefont{L.~N.} \bibnamefont{Bezmaternykh}},
  \bibnamefont{and} \bibinfo{author}{\bibfnamefont{V.~L.}
  \bibnamefont{Temerov}}, \bibinfo{journal}{Phys. Rev. B}
  \textbf{\bibinfo{volume}{75}}, \bibinfo{pages}{224435}
  (\bibinfo{year}{2007}),
  \urlprefix\url{https://link.aps.org/doi/10.1103/PhysRevB.75.224435}.

\bibitem[{\citenamefont{Chukalina et~al.}(2010)\citenamefont{Chukalina, Popova,
  Bezmaternykh, and Gudim}}]{chukalina_pla_2010}
\bibinfo{author}{\bibfnamefont{E.~P.} \bibnamefont{Chukalina}},
  \bibinfo{author}{\bibfnamefont{M.~N.} \bibnamefont{Popova}},
  \bibinfo{author}{\bibfnamefont{L.~N.} \bibnamefont{Bezmaternykh}},
  \bibnamefont{and} \bibinfo{author}{\bibfnamefont{I.~A.} \bibnamefont{Gudim}},
  \bibinfo{journal}{Phys. Lett. A} \textbf{\bibinfo{volume}{374}},
  \bibinfo{pages}{1790} (\bibinfo{year}{2010}).

\bibitem[{\citenamefont{Volkov et~al.}(1985)\citenamefont{Volkov, Goncharov,
  Kozlov, Lebedev, and Prokhorov}}]{volkov_infrared_1985}
\bibinfo{author}{\bibfnamefont{A.~A.} \bibnamefont{Volkov}},
  \bibinfo{author}{\bibfnamefont{Y.~G.} \bibnamefont{Goncharov}},
  \bibinfo{author}{\bibfnamefont{G.~V.} \bibnamefont{Kozlov}},
  \bibinfo{author}{\bibfnamefont{S.~P.} \bibnamefont{Lebedev}},
  \bibnamefont{and} \bibinfo{author}{\bibfnamefont{A.~M.}
  \bibnamefont{Prokhorov}}, \bibinfo{journal}{Infrared Phys.}
  \textbf{\bibinfo{volume}{25}}, \bibinfo{pages}{369} (\bibinfo{year}{1985}),
  \urlprefix\url{http://www.sciencedirect.com/science/article/B6X3W-46K4CH2-29/2/43ef2c1b497859011f2f67a144d4bb72}.

\bibitem[{\citenamefont{Kuzmenko et~al.}(2016)\citenamefont{Kuzmenko, Mukhin,
  Ivanov, Komandin, Shuvaev, Pimenov, Dziom, Bezmaternykh, and
  Gudim}}]{kuzmenko_prb_2016}
\bibinfo{author}{\bibfnamefont{A.~M.} \bibnamefont{Kuzmenko}},
  \bibinfo{author}{\bibfnamefont{A.~A.} \bibnamefont{Mukhin}},
  \bibinfo{author}{\bibfnamefont{V.~Y.} \bibnamefont{Ivanov}},
  \bibinfo{author}{\bibfnamefont{G.~A.} \bibnamefont{Komandin}},
  \bibinfo{author}{\bibfnamefont{A.}~\bibnamefont{Shuvaev}},
  \bibinfo{author}{\bibfnamefont{A.}~\bibnamefont{Pimenov}},
  \bibinfo{author}{\bibfnamefont{V.}~\bibnamefont{Dziom}},
  \bibinfo{author}{\bibfnamefont{L.~N.} \bibnamefont{Bezmaternykh}},
  \bibnamefont{and} \bibinfo{author}{\bibfnamefont{I.~A.} \bibnamefont{Gudim}},
  \bibinfo{journal}{Phys. Rev. B} \textbf{\bibinfo{volume}{94}},
  \bibinfo{pages}{174419} (\bibinfo{year}{2016}),
  \urlprefix\url{https://link.aps.org/doi/10.1103/PhysRevB.94.174419}.

\bibitem[{\citenamefont{Gudim et~al.}(2008)\citenamefont{Gudim, Pankrats,
  Durna{\u{\i}}kin, Petrakovski{\u{\i}}, Bezmaternykh, Szymczak, and
  Baran}}]{gudim_crr_2008}
\bibinfo{author}{\bibfnamefont{I.~A.} \bibnamefont{Gudim}},
  \bibinfo{author}{\bibfnamefont{A.~I.} \bibnamefont{Pankrats}},
  \bibinfo{author}{\bibfnamefont{E.~I.} \bibnamefont{Durna{\u{\i}}kin}},
  \bibinfo{author}{\bibfnamefont{G.~A.} \bibnamefont{Petrakovski{\u{\i}}}},
  \bibinfo{author}{\bibfnamefont{L.~N.} \bibnamefont{Bezmaternykh}},
  \bibinfo{author}{\bibfnamefont{R.}~\bibnamefont{Szymczak}}, \bibnamefont{and}
  \bibinfo{author}{\bibfnamefont{M.}~\bibnamefont{Baran}},
  \bibinfo{journal}{Crystallogr. Rep.} \textbf{\bibinfo{volume}{53}},
  \bibinfo{pages}{1140} (\bibinfo{year}{2008}),
  \urlprefix\url{http://dx.doi.org/10.1134/S1063774508070080}.

\bibitem[{\citenamefont{Demidov et~al.}(2013)\citenamefont{Demidov, Volkov,
  Gudim, Eremein, and Temerov}}]{demidov_jetp_2013}
\bibinfo{author}{\bibfnamefont{A.~A.} \bibnamefont{Demidov}},
  \bibinfo{author}{\bibfnamefont{D.~V.} \bibnamefont{Volkov}},
  \bibinfo{author}{\bibfnamefont{I.~A.} \bibnamefont{Gudim}},
  \bibinfo{author}{\bibfnamefont{E.~V.} \bibnamefont{Eremein}},
  \bibnamefont{and} \bibinfo{author}{\bibfnamefont{V.~L.}
  \bibnamefont{Temerov}}, \bibinfo{journal}{J. Exp. Theor. Phys.}
  \textbf{\bibinfo{volume}{116}}, \bibinfo{pages}{800} (\bibinfo{year}{2013}),
  \urlprefix\url{http://dx.doi.org/10.1134/S1063776113050038}.

\bibitem[{\citenamefont{Kadomtseva et~al.}(2006)\citenamefont{Kadomtseva,
  Krotov, Popov, and Vorob'ev}}]{kadomtseva_ltp_2006}
\bibinfo{author}{\bibfnamefont{A.~M.} \bibnamefont{Kadomtseva}},
  \bibinfo{author}{\bibfnamefont{S.~S.} \bibnamefont{Krotov}},
  \bibinfo{author}{\bibfnamefont{Y.~F.} \bibnamefont{Popov}}, \bibnamefont{and}
  \bibinfo{author}{\bibfnamefont{G.~P.} \bibnamefont{Vorob'ev}},
  \bibinfo{journal}{Low Temp. Phys.} \textbf{\bibinfo{volume}{32}},
  \bibinfo{pages}{709} (\bibinfo{year}{2006}),
  \urlprefix\url{http://scitation.aip.org/content/aip/journal/ltp/32/8/10.1063/1.2219494}.

\end{thebibliography}

\end{document}